\documentclass[12pt]{article}
\usepackage{makeidx}
\usepackage{amssymb}
\usepackage{amsfonts}
\usepackage{amsmath}
\usepackage{graphicx}
\usepackage{setspace}
\usepackage{authblk}
\usepackage{units}
\usepackage[left=1.6cm,top=2cm,right=1.6cm]{geometry}

\begin{document}

\title{\textbf{Magnetic response to applied electrostatic field in external magnetic
field}}
\author[1,2]{T. C. Adorno\thanks{tadorno@usp.br, tadorno@ufl.edu}}
\author[1]{D. M. Gitman\thanks{gitman@dfn.if.usp.br}}
\author[3]{A. E. Shabad\thanks{shabad@lpi.ru}}
\affil[1]{\textit{Instituto de F\'{\i}sica, Universidade de S\~{a}o Paulo, Caixa Postal 66318, CEP 05508-090, S\~{a}o Paulo, S.P., Brazil;}}
\affil[2]{\textit{Department of Physics, University of Florida, 2001 Museum Road, Gainesville, FL 32611-8440, USA;}}
\affil[3]{\textit{P. N. Lebedev Physics Institute, 117924 Moscow, Russia.}}

\maketitle

\onehalfspacing

\begin{abstract}
We show, within QED and other possible nonlinear theories, that a static
charge localized in a finite domain of space becomes a magnetic dipole, if
it is placed in an external (constant and homogeneous) magnetic field in the
vacuum. The magnetic moment is quadratic in the charge, depends on its size
and is parallel to the external field, provided the charge distribution is
at least cylindrically symmetric. This magneto-electric effect is a
nonlinear response of the magnetized vacuum to an applied electrostatic
field. Referring to a simple example of a spherically-symmetric applied
field, the nonlinearly induced current and its magnetic field are found
explicitly throughout the space, the pattern of lines of force is depicted,
both inside and outside the charge, which resembles that of a standard
solenoid of classical magnetostatics.
\end{abstract}

\newpage

\section{Introduction}

\bigskip

With the two recent papers \cite{ShaGit2012}, \cite{Caio} we started a
series of works aimed at studying quantum electrodynamics (QED) (as well as
other nonlinear Abelian theories that may be historically traced back to 
\cite{Mie12}) under the conditions where the intrinsic nonlinearity of the
theory shows itself not only as interaction of electromagnetic fields with a
strong background, but also with themselves.

Manifestations of nonlinearity of the first type mentioned have been a focus
of attention during many years since, perhaps, the pioneering works \cite%
{Erber66BB70, Birula} (see \cite{Dittrich} and more recent papers \cite%
{Karbstein, ShaUso10} for some reviews of the subsequent advances in that
field). The strong background was served in the corresponding studies by the
constant and homogeneous electromagnetic field (note, however, Ref. \cite%
{Gies}, where a certain inhomogeneity was introduced, and Ref. \cite%
{GavGit12}, where it was shown that a strong nonhomogeneous magnetic field
is able to produce pairs of neutral fermions from the vacuum) and by the
field of a plane wave (see the review of the laser-associated researches in 
\cite{laser}), because in these cases the influence of the background could
be exactly taken into account for arbitrarily large value of their amplitude
through the use of exact solutions of the Dirac equation available for such
cases. The Dirac propagators for the virtual electrons and positrons in
Feynman graphs for the vacuum polarization were the agents of interaction
with the background field in the intermediate state.

In those works the varying electromagnetic fields are treated as small
perturbations of the background, and only the effects linear in their
amplitudes are taken into account, such as birefringence, photon capture by
a magnetic field \cite{ShaUso82, ShaUso85}, modification of the Coulomb law 
\cite{shabus, SadSod07, MacVys11}, magnetic shift of the critical charge
value \cite{GodMacVys12}, -- and even the positronium collapse \cite%
{ShaUso06} may be placed among effects of this class. The arena of
applicability of these results is mostly the pulsars and magnetars,
possessing sufficiently large magnetic fields.

In contrast with the above, in Refs. \cite{ShaGit2012}, \cite{Caio} and in
the present paper we are considering the effects, quadratic and cubic in the
amplitude of the perturbation. As a matter of fact, two important \ special
examples of such effects were studied before, which were the processes of
photon splitting in a magnetic \cite{Birula, Adler, Stoneham, Split} and
crossed \cite{Papanyan} fields, and the light-by-light scattering \ \cite%
{BerLifPit}, all taken on the photon mass shell. Our goal is to deal with
excitations of the vacuum, different from photons, subject to nonlinear
version of the Maxwell equations, stemming from QED or, more generally,
intrinsic to any other nonlinear electrodynamics. Whereas handling
many-photon matrix elements beyond the photon mass shell, necessary for
addressing a general nonlinear problem, turns out to be overcomplicated, we
succeeded to indicate a simple approximation able to take responsibility for
nonlinear effects in a universal manner and independent of an expansion in
powers of the fine structure constant $\alpha ,$ as far as QED is concerned.
All kernels in the nonlinear integro-differential Maxwell equations are
given in terms of variational field-derivatives of the effective action $%
\Gamma ,$ defined as \cite{weinberg} the generating functional of
irreducible vertices in QED, or as an action that fixes a theory in other
versions of nonlinear electrodynamics. The approximation we are dealing with
is referred to as the local or infrared approximation. It assumes that the
effective action functional is local, i.e., it does not depend on the
space-time derivatives of the field strength. True, this assumption
restricts the range of applicability to only slow-varying fields in space
and time, but it enables us to efficiently advance in describing the
effects, cubic and quadratic in the field strength. By acting along these
lines where there is no background \ field \cite{Caio}, we reproduced the
known \cite{WichKroll, BerLifPit} correction to the Coulomb field which is
cubic in the charge that produces it, and found cubic equations for dipole
moments of selfinteracting fields of magnetic and electric dipoles that may
be also viewed upon as nonlinear renormalizations of these quantities.\ In
Ref. \cite{ShaGit2012} we studied quadratic response of the background
constant and homogeneous magnetic field to an applied electric field of a
static charge at rest, and we found this response to be purely magnetic. In
the present paper we continue the investigation of that magneto-electric
effect, and we establish that the static charge placed into a background
magnetic field is a magnetic dipole with its magnetic moment proportional to
the charge squared and parallel to the background field, unless the charge
distribution violates the initial cylindric symmetry of the problem. (This
situation resembles our results \cite{AdoShaGitVas11} in noncommutative
electrodynamics.) More explicit formulae for the magnetic field in the short
and long ranges, for its lines of force and for the magnetic moment are
presented referring to a simple example, where the applied electrostatic
field is central-symmetric and would correspond, if in a nondispersive
vacuum, to a charge distributed homogeneously inside a finite-radius sphere.

In the rest of the present Introduction we recall the basic equations of
nonlinear electrodynamics, truncated at the third power of the varying
field, against a constant homogeneous background, define the kernels in them
as the second- and the third-rank polarization tensors in terms of the
field-derivatives of the effective action, introduce nonlinearly-induced
currents, and give expressions for the varying fields in terms of the
currents based on the use of eigenvector expansion of the second-rank
polarization tensor and the photon Green function. In Section 2 the second-
and third-rank polarization tensors in a magnetic field are written in the
infrared approximation as expressed in terms of the second and third
derivatives of the delta-functions of the coordinate differences. The
previously found linear-response correction to the Coulomb field of an
arbitrarily distributed static charge at large distances in a magnetic field
are given for completeness, and the general structure of quadratic response,
which is purely magnetic, is analyzed in terms of eigenmodes contribution.
In Section 3 very explicit expressions for the induced current and the
magnetic field are fully elaborated using the simplest example, where the
applied electric field is that of a homogeneously charged sphere of finite
radius. Differential equations for the shape of the magnetic lines of force
are solved, and the resulting pattern is drawn in Figs. 1 and 2 inside and
outside the charge. The magnetic moment of the charge, proportional to its
square, is given in terms of the background magnetic field and the radius of
the charge. In Section 4 it is shown that spherically-nonsymmetric charge
distributions also are characterized by a magnetic dipole moment, which is
parallel to the background magnetic field, if the charge distribution does
not specify any new direction in the space.

\bigskip

\bigskip

\subsection{Nonlinear Maxwell equations}

The exact electromagnetic field equations of QED with an external 4-current $%
j_{\mu }$ are the Euler-Lagrange equations $\frac{\delta S_{\mathrm{tot}}%
\left[ A\right] }{\delta A^{\rho }(x)}=0$ that originate from the total
action $S_{\mathrm{tot}}\left[ A\right] $ \footnote{%
Greek indices span the 4-dimensional Minkowski space-time taking the values
0,1,2,3. The metric tensor is $\eta _{\mu \nu }=\mathrm{diag}(-1,+1,+1,+1)$
and bold symbols are reserved for three-dimensional Euclidean vectors (for
instance $\mathbf{A}(x)=\left( A^{i}(x)=A_{i}(x)\right) \,,\,\,i=1,2,3$. The
Heaviside-Lorentz system of units is used throughout the paper.}%
\begin{eqnarray}
&&S_{\mathrm{tot}}\left[ A\right] =S_{\mathrm{Max}}\left[ A\right] +\Gamma %
\left[ A\right] +S_{\mathrm{int}}\left[ A\right] \,,\ \ S_{\mathrm{Max}}%
\left[ A\right] =-\int \mathfrak{F}\left( x\right) d^{4}x\,,  \label{01} \\
&&\Gamma \left[ A\right] =\int \mathcal{L}\left( x\right) d^{4}x\,,\ \ S_{%
\mathrm{int}}\left[ A\right] =-\int j_{\mu }\left( x\right) A^{\mu }\left(
x\right) d^{4}x\,,\ \   \notag \\
&&\mathfrak{F}\left( x\right) =\frac{1}{4}F^{\mu \nu }F_{\mu \nu },\text{ }%
F^{\mu \nu }=\partial ^{\mu }A^{\nu }(x)-\partial ^{\nu }A^{\mu }(x),\text{ }%
\partial ^{\mu }=\frac{\partial }{\partial x_{\mu }},  \notag
\end{eqnarray}%
where $S_{\mathrm{Max}}\left[ A\right] $ is the free Maxwell action, $\Gamma %
\left[ A\right] $ is the effective action, and $\mathcal{L}\left( x\right) $
is the effective Lagrangian. Under the effective action $\Gamma $ we
understand in QED the generating functional of one-particle-irreducible
vertices \cite{weinberg}. Alternatively, it may be any action defining a
nonlinear electrodynamics other than QED. Due to the gauge invariance it, as
a matter of fact, depends only on the field strengths $F^{\mu \nu },$ and
not of the 4-vector potentials $A^{\nu }.$\ Besides, only the relativistic
invariant combinations $\mathfrak{F}\left( z\right) =\frac{1}{4}F^{\mu \nu
}F_{\mu \nu }$ and $\mathfrak{G}=\frac{1}{4}F^{\rho \sigma }\tilde{F}_{\rho
\sigma }\mathbf{,}$ where the dual field tensor is defined as $\tilde{F}%
_{\rho \sigma }=\frac{1}{2}\epsilon _{\rho \sigma \lambda \kappa }F^{\lambda
\kappa },$ of the field strengths make the arguments of $\Gamma $ and $%
\mathcal{L}$. In QED the effective action contains the exhausting and final
information of the theory in the photon sector, and is subject to
calculation within one or another dynamic scheme or approximation,
especially the perturbation theory.

Expanding (\ref{01}) in power series of the small electromagnetic field $%
a_{\mu }\left( x\right) =A_{\mu }(x)-\mathcal{A}_{\mu }(x)$ above the
external background $\mathcal{A}(x)$ of a constant and homogeneous magnetic
field $\mathbf{B=}$ $\left( \boldsymbol{\nabla }\times \boldsymbol{\mathcal{A%
}}\right) $, and restricting ourselves to the next-to-leading term, the
minimum action condition becomes the nonlinear Maxwell equations:%
\begin{eqnarray}
&&\left[ \square \eta _{\rho \nu }-\partial _{\rho }\partial _{\nu }\right]
a^{\nu }\left( x\right) +\int d^{4}x^{\prime }\Pi _{\alpha \rho }\left(
x^{\prime },x\right) a^{\alpha }\left( x^{\prime }\right)  \notag \\
&&+\frac{1}{2}\int d^{4}x^{\prime }d^{4}x^{\prime \prime }\Pi _{\alpha \beta
\rho }\left( x^{\prime },x^{\prime \prime },x\right) a^{\alpha }\left(
x^{\prime }\right) a^{\beta }\left( x^{\prime \prime }\right) =j_{\rho
}\left( x\right) \,,  \label{ME} \\
&&\Pi ^{\alpha \rho }\left( x^{\prime },x\right) =\left. \frac{\delta
^{2}\Gamma }{\delta A_{\alpha }\left( x^{\prime }\right) \delta A_{\rho
}\left( x\right) }\right\vert _{A=\mathcal{A}}\,,  \notag \\
&&\Pi ^{\alpha \beta \rho }\left( x^{\prime },x^{\prime \prime },x\right)
=\left. \frac{\delta ^{3}\Gamma }{\delta A_{\alpha }\left( x^{\prime
}\right) \delta A_{\beta }\left( x^{\prime \prime }\right) \delta A_{\rho
}\left( x\right) }\right\vert _{A=\mathcal{A}}\,,  \label{02}
\end{eqnarray}%
where $\Pi ^{\rho \alpha }\left( x^{\prime },x\right) $, $\Pi ^{\alpha \beta
\rho }\left( x^{\prime },x^{\prime \prime },x\right) $ are the second- and
third-rank polarization tensor in an external magnetic field, respectively.
Note that as long as the external field is constant in space-time, the
polarization tensors (\ref{02}) are functions on the differences of their
arguments. In obtaining Eq. (\ref{ME}) the zero-order power of the field $%
a_{\mu }\left( x\right) $ does not appear, since the space-time-independent
external field $\mathcal{F}^{\mu \nu }=\partial ^{\mu }\mathcal{A}^{\nu
}-\partial ^{\nu }\mathcal{A}^{\mu }$ exactly obeys the sourceless Maxwell
equations $\partial _{\mu }\mathcal{F}^{\mu \nu }=\left. \frac{\delta \Gamma 
}{\delta A_{\nu }\left( x\right) }\right\vert _{A=\mathcal{A}}=0$\emph{\ . }%
The\emph{\ }power series has been truncated to the next-to-leading
correction (i.e., we have neglected $\sim O\left( a^{3}\right) $).

Defining the \textit{nonlinear current} as%
\begin{equation}
j_{\rho }^{\mathrm{nl}}\left( x\right) =-\frac{1}{2}\int d^{4}x^{\prime
}d^{4}x^{\prime \prime }\Pi _{\nu \sigma \rho }\left( x^{\prime },x^{\prime
\prime },x\right) a^{\nu }\left( x^{\prime }\right) a^{\sigma }\left(
x^{\prime \prime }\right) \,,\,  \label{nonlincurr}
\end{equation}%
we may write (\ref{ME}) in the following way%
\begin{equation}
\left[ \square \eta _{\rho \nu }-\partial _{\rho }\partial _{\nu }\right]
a^{\nu }\left( x\right) +\int d^{4}x^{\prime }\Pi _{\rho \alpha }\left(
x^{\prime },x\right) a^{\alpha }\left( x^{\prime }\right) =j_{\rho }\left(
x\right) +j_{\rho }^{\mathrm{nl}}\left( x\right) \,.  \label{aa}
\end{equation}

While solving the nonlinear set of Maxwell equations (\ref{aa}), (\ref%
{nonlincurr}) we should not, strictly speaking, exceed the initial accuracy.
This implies that we treat the nonlinearity iteratively. To this end we
divide its solution into two parts as $a_{\nu }(x)=a_{\nu }^{\text{\textrm{%
lin}}}(x)+a_{\nu }^{\text{\textrm{nl}}}(x),$ with $a_{\nu }^{\text{\textrm{%
lin}}}(x)\gg a_{\nu }^{\text{\textrm{nl}}}(x).$ Then, defining the linear
field $a_{\nu }^{\text{\textrm{lin}}}(x)$\ as a solution to the equation%
\begin{equation}
\left[ \square \eta _{\rho \nu }-\partial_{\rho }\partial _{\nu }\right] a_{%
\text{\textrm{lin}}}^{\nu }\left( x\right) +\int d^{4}x^{\prime }\Pi
_{\rho\alpha }\left( x^{\prime },x\right) a^{\alpha }_{\text{\textrm{lin}}%
}\left( x^{\prime }\right) =j_{\rho }\left( x\right) ,  \label{alin}
\end{equation}%
we get, as the first iteration, that the nonlinear correction $a_{\nu }^{%
\text{nl}}(x)$ to it is subject to the linear inhomogeneous equation 
\begin{equation}
\left[ \square \eta _{\rho \nu }-\partial _{\rho }\partial _{\nu }\right] a_{%
\mathrm{nl}}^{\nu }\left( x\right) +\int d^{4}x^{\prime }\Pi _{\rho \alpha
}\left( x^{\prime },x\right) a_{\text{\textrm{nl}}}^{\alpha }\left(
x^{\prime }\right) =\left. j_{\sigma }^{\text{nl}}\left( x^{\prime }\right)
\right\vert _{a=a_{\text{lin}}}\,,  \label{anonlin1}
\end{equation}%
where 
\begin{equation}
\left. j_{\rho }^{\text{nl}}\left( x^{\prime }\right) \right\vert _{a=a_{%
\text{lin}}}=-\frac{1}{2}\int d^{4}x^{\prime }d^{4}x^{\prime \prime }\Pi
_{\rho \nu \sigma }\left( x^{\prime },x^{\prime \prime },x\right) a_{\text{%
\textrm{lin}}}^{\nu }\left( x^{\prime }\right) a_{\text{\textrm{lin}}%
}^{\sigma }\left( x^{\prime \prime }\right)  \label{takencur}
\end{equation}%
is the nonlinear current (\ref{nonlincurr}) taken at the linear value (\ref%
{solut1}) of the field $a^{\nu }\left( x\right) =a_{\mathrm{lin}}^{\nu
}\left( x\right) .$The solution of Eqs. (\ref{alin}) and (\ref{anonlin1})
may be written as%
\begin{eqnarray}
a_{\mathrm{lin}}^{\nu }\left( x\right) &=&\int d^{4}x^{\prime }D^{\nu \sigma
}\left( x,x^{\prime }\right) j_{\sigma }\left( x^{\prime }\right) \,,
\label{solut1} \\
a_{\mathrm{nl}}^{\nu }\left( x\right) &=&\int d^{4}x^{\prime }D^{\nu \sigma
}\left( x,x^{\prime }\right) \left. j_{\sigma }^{\text{nl}}\left( x^{\prime
}\right) \right\vert _{a=a_{\text{lin}}},  \label{solut2}
\end{eqnarray}
The photon Green function in $D_{\mu \nu }(x,x^{\prime })$ above is defined
as the inverse operator: 
\begin{equation}
D_{\mu \nu }^{-1}(x-x^{\prime })=\left[ \eta _{\mu \nu }\square -\partial
_{\mu }\partial _{\nu }\right] \delta ^{(4)}(x^{\prime }-x)+\Pi _{\mu \nu
}(x-x^{\prime }).  \label{propagator}
\end{equation}%
Its Fourier transform $D^{\nu }{}_{\rho }(k)$\ with respect to the
coordinate difference should satisfy the following algebraic inhomogeneous
equation%
\begin{equation}
\left[ k^{2}\eta _{\mu \nu }-k_{\mu }k_{\nu }-\Pi _{\mu \nu }(k)\right]
D^{\nu }{}_{\rho }(k)=(\eta _{\mu \rho }-\frac{k_{\mu }k_{\rho }}{k^{2}}).
\label{propeq}
\end{equation}%
To solve this equation it is convenient to use the diagonal representation
for the second-rank polarization operator in a magnetic field \cite%
{Shabad72-75}%
\begin{equation}
\Pi _{\mu \tau }(k,p)=\delta (k-p)\Pi _{\mu \tau }(k)\,,\ \ \Pi _{\mu \tau
}(k)=\sum_{c=1}^{3}\varkappa _{c}(k)\frac{\flat _{\mu }^{(c)}\flat _{\tau
}^{(c)}}{(\flat ^{(c)})^{2}}  \label{diag}
\end{equation}%
in terms of the mutually orthogonal 4-vectors $\flat _{\mu }^{(c)}$%
\begin{equation}
\flat _{\mu }^{(1)}=(\mathcal{F}^{2}k)_{\mu }k^{2}-k_{\mu }(k\mathcal{F}%
^{2}k)\,,\ \ \flat _{\mu }^{(2)}=(\mathcal{\tilde{F}}k)_{\mu }\,,\ \ \flat
_{\mu }^{(3)}=(\mathcal{F}k)_{\mu }\,,\ \ \flat _{\mu }^{(4)}=k_{\mu }\,,
\label{eigenvect}
\end{equation}%
where $(\mathcal{\tilde{F}}k)_{\mu }\equiv \mathcal{\tilde{F}}_{\mu \tau
}k^{\tau }$, $(\mathcal{F}k)_{\mu }\equiv \mathcal{F}_{\mu \tau }k^{\tau }$, 
$(\mathcal{F}^{2}k)_{\mu }\equiv \mathcal{F}_{\mu \tau }^{2}k^{\tau }$, $k%
\mathcal{F}^{2}k\equiv k^{\mu }\mathcal{F}_{\mu \tau }^{2}k^{\tau }$, which
are the eigenvectors of the polarization operator%
\begin{equation}
\Pi _{\mu \tau }\flat ^{(c)\tau }=\varkappa _{c}(k)\flat _{\mu }^{(c)},
\label{eigeq}
\end{equation}%
the scalar functions $\varkappa _{c}(k)$\ being its four eigenvalues, $%
\varkappa _{4}(k)=0$. The eigenvalues $\varkappa _{c}(k)$ depend on $%
\mathfrak{F}$ and on any two of the three momentum-containing Lorentz
invariants $k^{2}=\mathbf{k}^{2}-k_{0}^{2},\;k\mathcal{F}^{2}k,\;k\mathcal{%
\tilde{F}}^{2}k$ , subject to one relation $\frac{k\mathcal{\tilde{F}}^{2}k}{%
2\mathfrak{F}}-k^{2}=\frac{k\mathcal{F}^{2}k}{2\mathfrak{F}}$, where $%
\mathfrak{F}$ is taken on the external field, $2\mathfrak{F=}$ $B^{2}.$ The
solution of (\ref{propeq}) has arbitrary longitudinal part:

\begin{align}
D_{\mu \tau }(k)& =\sum_{c=1}^{4}D^{(c)}(k)\frac{\flat _{\mu }^{(c)}\flat
_{\tau }^{(c)}}{(\flat ^{(c)})^{2}},  \notag \\
D^{(c)}(k)& =\left\{ 
\begin{tabular}{cc}
$(k^{2}-\varkappa _{c}(k))^{-1},$ & $c=1,2,3$ \\ 
\textrm{arbitrary}, & $c=4$%
\end{tabular}%
\,.\right.  \label{D}
\end{align}%
It also has a diagonal form in the same terms as (\ref{diag}). This
propagator has three components, corresponding to separate eigenmodes. Each
of them has a pole in the 4-momentum plane, where solutions of the
corresponding dispersion equations lie, i.e. on the photon mass shell,
defined by the equations $k^{2}-\varkappa _{c}(k)=0$.

Now, solutions (\ref{solut1}), (\ref{solut2}) may be, respectively, written
as%
\begin{eqnarray}
a_{\mu }^{\mathrm{lin}}(k) &=&\sum_{c=1}^{4}\frac{1}{(k^{2}-\varkappa
_{c}(k))}\frac{\flat _{\mu }^{(c)}}{(\flat ^{(c)})^{2}}(j^{\tau }\left(
k\right) \flat _{\tau }^{(c)})\,,  \label{potmag1} \\
a_{\mu }^{\mathrm{nl}}(k) &=&\sum_{c=1}^{4}\frac{1}{(k^{2}-\varkappa _{c}(k))%
}\frac{\flat _{\mu }^{(c)}}{(\flat ^{(c)})^{2}}(j_{\mathrm{nl}}^{\tau
}\left( k\right) \flat _{\tau }^{(c)})\,.  \label{potmag2}
\end{eqnarray}

\section{Response of magnetized vacuum to static electric field in the
infrared approximation}

In the rest of the paper we shall be treating equations of Subsection 2.1 in
the low-momentum-low-frequency (infrared) approximation, $k_{\mu }\sim 0,$
which is stemming from the assumption that the effective action $\Gamma
\lbrack A]$ is a local functional of the field strengths $F_{\mu \nu }$ in
the sense that it does not contain their space- and time-derivatives.
Examples of such action are the Heisenberg-Euler action available in the
one-loop \cite{BerLifPit} and two-loop \cite{Ritus2loop} approximations in
QED, the Born-Infeld \cite{BorInf34} action \textit{etc}. Within the local
limit the second- and third-rank polarization tensors (\ref{02}) were
calculated in \cite{ShaGit2012} to give the result \footnote{%
The fourth-rank tensor in the same approximation is also available \cite%
{Caio}}%
\begin{eqnarray}
\Pi _{\mu \nu }\left( x,x^{\prime }\right) &=&\left\{ \mathcal{L}_{\mathfrak{%
F}}\left( \frac{\partial ^{2}}{\partial x^{\mu }\partial x^{\nu }}-\eta
_{\mu \nu }\square ^{x}\right) \right.  \notag \\
&-&\left. \left( \mathcal{L}_{\mathfrak{FF}}\mathcal{F}_{\mu \alpha }%
\mathcal{F}_{\nu \beta }+\mathcal{L}_{\mathfrak{GG}}\mathcal{\tilde{F}}_{\mu
\alpha }\mathcal{\tilde{F}}_{\nu \beta }\right) \frac{\partial ^{2}}{%
\partial x_{\alpha }\partial x_{\beta }}\right\} \delta ^{\left( 4\right)
}\left( x-x^{\prime }\right) \,,  \label{07a} \\
\Pi _{\nu \rho \sigma }\left( x,x^{\prime },x^{\prime \prime }\right) &=&-%
\mathcal{O}_{\nu \rho \sigma \alpha \beta \gamma }\frac{\partial }{\partial
x_{\alpha }}\left( \frac{\partial }{\partial x_{\beta }}\delta ^{\left(
4\right) }\left( x-x^{\prime }\right) \right) \left( \frac{\partial }{%
\partial x_{\gamma }}\delta ^{\left( 4\right) }\left( x^{\prime }-x^{\prime
\prime }\right) \right) \,,  \label{07}
\end{eqnarray}%
where 
\begin{align}
& \mathcal{O}_{\mu \tau \sigma \alpha \beta \gamma }=\mathfrak{L_{GG}}\left[ 
\widetilde{\mathcal{F}}_{\gamma \sigma }\epsilon _{\alpha \mu \beta \tau }+%
\widetilde{\mathcal{F}}_{\alpha \mu }\epsilon _{\beta \tau \gamma \sigma }+%
\widetilde{\mathcal{F}}_{\beta \tau }\epsilon _{\alpha \mu \gamma \sigma }%
\right]  \notag \\
& +\mathfrak{L_{FF}}\left[ \left( \eta _{\mu \tau }\eta _{\alpha \beta
}-\eta _{\mu \beta }\eta _{\alpha \tau }\right) \mathcal{F}_{\gamma \sigma }+%
\mathcal{F}_{\alpha \mu }\left( \eta _{\tau \sigma }\eta _{\gamma \beta
}-\eta _{\beta \sigma }\eta _{\gamma \tau }\right) +\mathcal{F}_{\beta \tau
}\left( \eta _{\mu \sigma }\eta _{\gamma \alpha }-\eta _{\alpha \sigma }\eta
_{\gamma \mu }\right) \right]  \notag \\
& +\mathfrak{L_{FGG}}\left[ {\mathcal{F}}_{\alpha \mu \text{ }}\widetilde{%
\mathcal{F}}_{\beta \tau }\widetilde{\mathcal{F}}_{\gamma \sigma }+%
\widetilde{\mathcal{F}}_{\alpha \mu }\mathcal{F}_{\beta \tau }\widetilde{%
\mathcal{F}}_{\gamma \sigma }\text{\emph{+}}\widetilde{\mathcal{F}}_{\alpha
\mu }\widetilde{\mathcal{F}}_{\beta \tau }\mathcal{F}_{\gamma \sigma }\right]
\text{ }+\mathfrak{L_{FFF}}\mathcal{F}_{\alpha \mu }\mathcal{F}_{\beta \tau }%
\mathcal{F_{\gamma \sigma }\,},  \label{sixrank}
\end{align}%
which expresses them in terms of the derivatives of the effective Lagrangian
taken at the constant external field value, $\mathcal{F}^{\mu \nu }=\mathrm{%
const.}$,

\begin{align}
\mathcal{L}\mathfrak{_{F}}& =\left. \frac{d\mathcal{L}(\mathfrak{F},0))}{d%
\mathfrak{F}}\right\vert _{F=\mathcal{F}}\,,\ \ \mathcal{L}\mathfrak{_{FF}}%
=\left. \frac{d^{2}\mathcal{L}(\mathfrak{F},0)}{d\mathfrak{F}^{2}}%
\right\vert _{F=\mathcal{F}},\ \ \mathcal{L}\mathfrak{_{GG}}=\left. \frac{%
\partial ^{2}\mathcal{L}(\mathfrak{F},\mathfrak{G})}{\partial \mathfrak{G}%
^{2}}\right\vert _{F=\mathcal{F},\mathfrak{G}=0}\,,  \label{derivative1} \\
\mathcal{L}\mathfrak{_{FFF}}& \mathfrak{=}\left. \frac{d^{3}\mathcal{L}(%
\mathfrak{F},0)}{d\mathfrak{F}^{3}}\right\vert _{F=\mathcal{F}}\,,\ \ 
\mathcal{L}\mathfrak{_{FGG}}=\frac{d}{d\mathfrak{F}}\left. \frac{\partial
^{2}\mathcal{L}(\mathfrak{F},\mathfrak{G})}{\partial \mathfrak{G}^{2}}%
\right\vert _{F=\mathcal{F},\mathfrak{G}=0}\,.  \label{derivative two}
\end{align}%
It is taken into account that once the external field is purely magnetic in
a certain Lorentz frame, the invariant $\mathfrak{G}$ for it is zero while $%
\mathfrak{F}$ is positive.

It was established in \cite{Shabus2011} that the second-rank polarization
operator Eq. (\ref{07a}) has indeed the structure of (\ref{diag}) with the
egenvalues $\varkappa _{a}(k)$ in the infrared regime being%
\begin{eqnarray}
\left. \varkappa _{1}(k^{2},k\mathcal{F}^{2}k,\mathfrak{F})\right\vert
_{k\rightarrow 0} &=&k^{2}\mathcal{L}\mathfrak{_{F}},  \label{kappa1} \\
\left. \varkappa _{2}(k^{2},k\mathcal{F}^{2}k,\mathfrak{F})\right\vert
_{k\rightarrow 0} &=&k^{2}\mathcal{L}\mathfrak{_{F}}-(k\mathcal{\tilde{F}}%
^{2}k)\mathcal{L}\mathfrak{_{GG}},  \label{kappa2} \\
\left. \varkappa _{3}(k^{2},k\mathcal{F}^{2}k,\mathfrak{F})\right\vert
_{k\rightarrow 0} &=&k^{2}\mathcal{L}\mathfrak{_{F}}-(k\mathcal{F}^{2}k)%
\mathcal{L}\mathfrak{_{FF}}.  \label{kappa3}
\end{eqnarray}

Hitherto, we shall be dealing only with sources that are static in a
reference frame, where the external field is magnetic, i.e. time-independent
charges at rest in a magnetic field 
\begin{equation}
j_{\mu }(x)=\delta _{0\mu }q(\mathbf{x})\,,\ \ \text{$\tilde{j}$}_{\mu
}(k)=(2\pi)\delta _{0\mu }\delta (k_{0})\tilde{q}(\mathbf{k}),
\label{static}
\end{equation}%
where the tilde marks the Fourier-transformed function.

\subsection{Linear response -- modified Coulomb law at large distances}

Employing the source (\ref{static}) in (\ref{potmag1}) and taking into
account that at $k_{0}=0$ out of all the three (nontrivial) eigenvectors (%
\ref{eigenvect}) only $\flat _{\mu }^{(2)}$ has its zeroth component
different from zero, while its spatial components disappear $\flat
_{i}^{(2)}=0,$ we get 
\begin{equation}
a_{\text{\textrm{lin}}}^{0}(k)=(2\pi)\sum_{a=1}^{3}\frac{\delta (k_{0})%
\widetilde{q}(\mathbf{k})}{(k^{2}-\varkappa _{a}(k))}\frac{\flat _{0}^{(a)}}{%
(\flat ^{(a)})^{2}}\flat _{0}^{(a)}=\frac{(2\pi)\delta (k_{0})\widetilde{q}(%
\mathbf{k})}{(\mathbf{k}^{2}-\varkappa _{2}(\mathbf{k}))}\,,\ \ \text{$%
\boldsymbol{a}$}_{\text{\textrm{lin}}}(k)=0\,.  \label{linvecpot}
\end{equation}%
(We disregarded the longitudinal part $k^{\mu }$ that does not contribute to
the field strength). So, naturally, only static electric field is produced
by a static source at the linear level (\ref{solut1}), and no magnetic field.

Eq. (\ref{linvecpot}) is approximation-independent. In the infrared limit,
Eq. (\ref{kappa2}) is to be used in it with $k_{0}=0$ and $(k\tilde{F}%
^{2}k)=2\mathfrak{F}k_{3}^{2}=B^{2}k_{3}^{2}$ in the special frame, when the
latter is so oriented that axis 3 coincide with the direction of the
external magnetic field $\mathbf{B},$ $B=B_{3}=\mathcal{F}_{12},$ $B_{1,2}=0$%
. Then, in the coordinate space the linear potential (\ref{linvecpot})
becomes%
\begin{equation}
a_{\text{\textrm{lin}}}^{0}(\mathbf{x})=\frac{1}{(2\pi)^3}\int \frac{%
\widetilde{q}(\mathbf{k})e^{i\mathbf{k}_{\perp }\cdot \mathbf{x}_{\perp
}}d^{2}k_{\perp }e^{ik_{3}\cdot x_{3}}dk_{3}}{\left( 1-\mathcal{L}_{%
\mathfrak{F}}+2\mathfrak{F}\mathcal{L}_{\mathfrak{GG}}\right) k_{3}^{2}+%
\mathbf{k}_{\perp }^{2}\left( 1-\mathcal{L}_{\mathfrak{F}}\right) }=\frac{1}{%
(2\pi)^3}\int \widetilde{q}(\mathbf{k})\frac{e^{i\mathbf{k}_{\perp }\cdot 
\mathbf{x}_{\perp }}d^{2}k_{\perp }e^{ik_{3}\cdot x_{3}}dk_{3}}{\varepsilon
_{\text{\textrm{long}}}k_{3}^{2}+\mathbf{k}_{\perp }^{2}\varepsilon _{\text{%
\textrm{tr}}}}\,.  \label{linpot1}
\end{equation}%
where we have used expressions for longitudinal and transverse dielectric
constants from \cite{Shabus2011} (see also \cite{Chavez})

\begin{equation*}
1-\mathcal{L}_{\mathfrak{F}}=\varepsilon _{\text{\textrm{tr}}},\text{ }1-%
\mathcal{L}_{\mathfrak{F}}+2\mathfrak{F}\mathcal{L}_{\mathfrak{GG}%
}=\varepsilon _{\text{\textrm{long}}}\,,
\end{equation*}%
where the field invariant\ $\mathfrak{F}$ is henceforward taken on the
external field, $2\mathfrak{F=}$ $B^{2},$ unlike its previous more general
definition in (\ref{01}). The subscript $\perp $ indicates projection onto
(1, 2)-plane. The behavior of (\ref{linpot1}) at large distances $|\mathbf{x}%
_{\perp }|\rightarrow \infty ,x_{3}\rightarrow \infty $ is 

\begin{equation}
a_{\text{\textrm{lin}}}^{0}(\mathbf{x})\sim \frac{Q}{(2\pi)^3}\int \frac{e^{i%
\mathbf{k}_{\perp }\cdot \mathbf{x}_{\perp }}d^{2}k_{\perp }e^{ik_{3}\cdot
x_{3}}dk_{3}}{\varepsilon _{\text{\textrm{long}}}k_{3}^{2}+\mathbf{k}_{\perp
}^{2}\varepsilon _{\text{\textrm{tr}}}}\,,  \label{linpotass}
\end{equation}%
where $Q=\widetilde{q}(0)=\int q(\mathbf{x)}d^{3}x$ is the total charge in
understanding that the charge density is either compactly distributed inside
a volume or decreases sufficiently fast outside it so that this integral
converge.

By making the change of variables $\varepsilon _{\text{\textrm{long}}%
}^{1/2}k_{3}=k_{3}^{\prime },$ $\mathbf{k}_{\perp }\varepsilon _{\text{%
\textrm{tr}}}^{1/2}=\mathbf{k}_{\perp }^{\prime }$ the integral (\ref%
{linpotass}) is transformed to

\begin{eqnarray}
a_{\text{\textrm{lin}}}^{0}(\mathbf{x}) &\sim &\frac{Q}{(2\pi)^3\varepsilon
_{\text{\textrm{long}}}^{1/2}\varepsilon _{\text{\textrm{tr}}}}\int \frac{%
e^{i\mathbf{k}^{\prime }\cdot \mathbf{x}^{\prime }}d^{3}k^{\prime }}{\mathbf{%
k}^{\prime 2}}=\frac{Q}{4\pi \varepsilon _{\text{\textrm{long}}%
}^{1/2}\varepsilon _{\text{\textrm{tr}}}}\frac{1}{|\mathbf{x}^{\prime }|} 
\notag \\
&=&\frac{1}{4\pi \varepsilon _{\text{\textrm{long}}}^{1/2}\varepsilon _{%
\text{\textrm{tr}}}}\frac{Q}{\left( \mathbf{x}_{\perp }^{2}\varepsilon _{%
\text{\textrm{tr}}}^{-1}+x_{3}^{2}\varepsilon _{\text{\textrm{long}}%
}^{-1}\right) ^{1/2}}=\frac{1}{4\pi \varepsilon _{\text{\textrm{tr}}}^{1/2}}%
\frac{Q}{\left( \mathbf{x}_{\perp }^{2}\varepsilon _{\text{\textrm{long}}%
}+x_{3}^{2}\varepsilon _{\text{\textrm{tr}}}\right) ^{1/2}},  \label{Coulomb}
\end{eqnarray}%
where the notations $\mathbf{x}_{\perp }^{\prime }=\mathbf{x}_{\perp
}\varepsilon _{\text{\textrm{tr}}}^{-1/2},$ $x_{3}^{\prime
}=x_{3}\varepsilon _{\text{\textrm{long} }}^{-1/2}$ , $|\mathbf{x}^{\prime
}|=(|\mathbf{x}_{\perp }^{\prime }|^{2}+(x_{3}^{\prime })^{2})^{1/2}$ were
used. This anisotropic Coulomb law, resulted from the infrared limit (\ref%
{07a}) of the second-rank polarization operator, is thereby the
long-distance asymptotic behavior of the electrostatic potential produced by
a charge, locally distributed in space, in a constant magnetic field.

It may be also useful \ to write the scalar potential (\ref{Coulomb}) in an
O(3)-invariant way as a function of two rotational scalars:

\begin{equation}
a_{\text{\textrm{lin}}}^{0}(\mathbf{x})=a_{\text{\textrm{lin}}}^{0}(\mathbf{x%
}^{2},(\mathbf{B\cdot x}))=\frac{1}{4\pi \varepsilon _{\text{\textrm{tr}}%
}^{1/2}}\frac{Q}{\left( \mathbf{x}^{2}\varepsilon _{\text{\textrm{long}}}+(%
\mathbf{B\cdot x})^{2}\mathcal{L}_{\mathfrak{GG}}\right) ^{1/2}}.
\label{Coulomb1}
\end{equation}%
Previously \cite{shabus, SadSod07}, that potential was found in QED in the
whole space, the vicinity of the charge -- where the potential has the Debye
form -- included, \ starting from the off-shell calculations of the full
(free of the restrictive assumption $k_{\mu }$ $\sim 0)$ second-rank
off-shell polarization operator in a magnetic field, first performed within
the accuracy of one fermion loop in \cite{Shabad72-75}. If the one-loop
Heisenberg-Euler effective Lagrangian is taken for $\mathcal{L}$ in (\ref%
{Coulomb}), it makes (a corrected\footnote{%
The omission in \cite{shabus} was that \textbf{k}$_{\perp }$ was set equal
to zero when deriving Eq. (35) there. For large magnetic field in QED $%
\varepsilon _{\text{long}}$ grows linearly with the field, whereas $%
\varepsilon _{\text{tr}}$ remains $\approx 1,$ since $\mathcal{L}_{\mathfrak{%
F}}\sim \alpha \ln B/B_{\text{Sch}.}$ For this reason all conclusions made
in \cite{shabus} concerning the large-field behavior remain unaffected.}
form of) the large-distance behavior of the potential in QED.

\subsection{Quadratic response -- magneto-electric effect}

When the four-vector potential in expression (\ref{nonlincurr}) is chosen as 
$a^{\mu }\left( \mathbf{x}\right) \simeq a_{\text{lin}}^{\mu }\left( \mathbf{%
x}\right) =\delta _{0\mu }$ $a_{\text{lin}}^{0}\left( \mathbf{x}\right) ,$
so as to carry only electrostatic field $\mathbf{E\left( \mathbf{x}\right) =}%
-\mathbf{\nabla }a^{0}\left( \mathbf{x}\right) ,$ the nonlinear current (\ref%
{nonlincurr}), approximated as (\ref{takencur}) in accord with the iteration
(\ref{anonlin1}), (henceforth we omit the explicit indication that it is
taken on linear fields), becomes \cite{ShaGit2012}, up to the third and
fourth powers of the applied field%
\begin{eqnarray}
&&j_{\mathrm{nl}}^{0}\left( \mathbf{x}\right) =0\,,\ \ \mathbf{j}_{\mathrm{nl%
}}\left( \mathbf{x}\right) =\mathbf{j}_{\mathfrak{FF}}\left( \mathbf{x}%
\right) +\mathbf{j}_{\mathfrak{FGG}}\left( \mathbf{x}\right) +\mathbf{j}_{%
\mathfrak{GG}}\left( \mathbf{x}\right) \,,  \label{3} \\
&&\mathbf{j}_{\mathfrak{FF}}\left( \mathbf{x}\right) =\frac{\mathcal{L}_{%
\mathfrak{FF}}}{2}\left( \boldsymbol{\nabla }\times \mathbf{B}\right) 
\mathbf{E}^{2}\,,\ \ \mathbf{j}_{\mathfrak{FGG}}\left( \mathbf{x}\right) =-%
\frac{\mathcal{L}_{\mathfrak{FGG}}}{2}\left( \boldsymbol{\nabla }\times 
\mathbf{B}\right) \left( \mathbf{B}\cdot \mathbf{E}\right) ^{2}\,,  \notag \\
&&\mathbf{j}_{\mathfrak{GG}}\left( \mathbf{x}\right) =-\mathcal{L}_{%
\mathfrak{GG}}\left( \boldsymbol{\nabla }\times \mathbf{E}\right) \left( 
\mathbf{B}\cdot \mathbf{E}\right) \,,  \notag
\end{eqnarray}%
after the expression (\ref{07}) for the third-rank polarization tensor in
the local limit is used. Here every derivative acts on everything to the
right of it.

Note that in (\ref{3}), $\mathbf{E}$ is the applied electric field, the
nonlinear response to which is under consideration. Correspondingly, Eq. (%
\ref{3}) is quadratic with respect to $\mathbf{E.}$ On the contrary, the
external magnetic field $\mathbf{B=}$ $\left( \boldsymbol{\nabla }\times 
\boldsymbol{\mathcal{A}}\right) ,$ $B$ $=\left( 2\mathfrak{F}\right) ^{1/2%
\text{ }}$ enters (\ref{3}) with all powers, since the coefficients in it
depend on $B$ in a complicated way according to their definitions (\ref%
{derivative1}) and (\ref{derivative two}).

Let us discuss the structure of the nonlinear correction to the
electromagnetic field caused by the current (\ref{3}) following Eq. (\ref%
{solut2}). We appeal to representation (\ref{D}) for the photon propagator
in it. First we note that the mode $c=2$ does not contribute, since $j_{%
\mathrm{nl}}^{0}=0,$ and $\flat _{i}^{(2)}=(\mathcal{\tilde{F}}k)_{i}$ (\ref%
{eigenvect}) disappears, when multiplied by the Fourier transform $\mathbf{%
\widetilde{\mathbf{j}}}_{\mathrm{nl}}\left( \mathbf{k}\right) \delta (k_{0})$
of (\ref{3}). Also the zeroth components of the other two eigenvectors (\ref%
{eigenvect}) $\flat _{0}^{(1,3)}$\ vanish if taken at $k_{0}=0$. Thus we are
left with,%
\begin{equation}
a_{\mathrm{nl}}^{0}=0\,,\ \ a_{\text{\textrm{nl}}}^{i}(k)=\sum_{c=1,3}\frac{%
(2\pi) \delta (k_{0})}{(k^{2}-\varkappa _{c}(k))}\frac{\flat _{i}^{(c)}}{%
(\flat ^{(c)})^{2}}(\widetilde{j}_{j\mathrm{nl}}\left( \mathbf{k}\right)
\flat _{j}^{(c)})\,,  \label{1,3}
\end{equation}%
which indicates that the quadratic response to a static electric field is
purely magnetic.

It can be shown that Eqs. (\ref{1,3}) are in fact exact relations in the
electrostatic case independent of the infrared approximation. This implies
that in that instance only modes 1 and 3 propagate magnetic field. However,
in a spherically symmetric infrared example to be considered below it holds
that $\widetilde{j}_{j\mathrm{nl}}\left( \mathbf{k}\right) \flat
_{j}^{(1)}=0,$ so only the term $c=3$ contributes in the nonlinear
electromagnetic field (\ref{1,3}).

If the linear vacuum polarization (in fact, the magnetization) is neglected%
\footnote{%
The effect of the magnetization is of higher order in the fine-structure
constant. Its full account can be found in \cite{AdoGitSha2014}}, $\varkappa
_{1,3}=0,$ the nonlinear magnetic field $\mathbf{h}=\mathbf{k}\times 
\boldsymbol{a}_{\text{nl}}$ obtained from (\ref{1,3}) satisfies the standard
Maxwell equation $\mathbf{k}^{2}\boldsymbol{a}_{\text{nl}}=\mathbf{%
\widetilde{\mathbf{j}}}_{\mathrm{nl}}$. In this case this field follows from
(\ref{3}) to be \cite{ShaGit2012},%
\begin{equation}
h_{i}\left( \mathbf{x}\right) =\mathfrak{h}_{i}\left( \mathbf{x}\right) +%
\frac{\partial _{i}\partial _{k}}{4\pi }\mathfrak{I}_{k}\left( \mathbf{x}%
\right) \,,\ \ \mathfrak{I}_{k}\left( \mathbf{x}\right) =\int d^{3}y\frac{%
\mathfrak{h}_{k}\left( \mathbf{y}\right) }{\left\vert \mathbf{x}-\mathbf{y}%
\right\vert }\,,  \label{magfield}
\end{equation}%
where%
\begin{eqnarray}
&&\mathfrak{h}_{i}\left( \mathbf{x}\right) =\mathfrak{h}_{i}^{\mathfrak{FF}%
}\left( \mathbf{x}\right) +\mathfrak{h}_{i}^{\mathfrak{FGG}}\left( \mathbf{x}%
\right) +\mathfrak{h}_{i}^{\mathfrak{GG}}\left( \mathbf{x}\right) \,,  \notag
\\
&&\mathfrak{h}_{i}^{\mathfrak{FF}}=\frac{B_{i}}{2}\mathcal{L}_{\mathfrak{FF}}%
\mathbf{E}^{2}\,,\ \mathfrak{h}_{i}^{\mathfrak{FGG}}=-\frac{B_{i}}{2}%
\mathcal{L}_{\mathfrak{FGG}}\left( \mathbf{B}\cdot \mathbf{E}\right)
^{2}\,,\ \ \mathfrak{h}_{i}^{\mathfrak{GG}}=-\mathcal{L}_{\mathfrak{GG}%
}\left( \mathbf{B}\cdot \mathbf{E}\right) E_{i}\,.  \label{hgerman}
\end{eqnarray}

\section{Quadratic magnetic response to a spherically-symmetric applied
electric field. Simple example}

In this Section, in order to present the magneto-electric effect in its most
explicit way, we shall consider the magnetic field, which is the response of
the vacuum to the applied electric field, whose vector potential is chosen
to be the following smooth spheric-symmetrical Coulomb-like function%
\begin{eqnarray}
&&a_{0}\left( r\right) =a_{0}^{\mathrm{I}}\left( r\right) \theta \left(
R-r\right) +a_{0}^{\mathrm{II}}\left( r\right) \theta \left( r-R\right)
\,,\,\,r=\left\vert \mathbf{x}\right\vert  \label{1.1} \\
&&a_{0}^{\mathrm{I}}\left( r\right) =-\frac{Ze}{8\pi R^{3}}r^{2}+\frac{3Ze}{%
8\pi R}\,,\ \ a_{0}^{\mathrm{II}}\left( r\right) =\frac{Ze}{4\pi r}\,. 
\notag
\end{eqnarray}%
Here $\theta \left( z\right) $ is the Heaviside unit step function, defined
as

\begin{equation*}
\theta \left( z\right) =\left\{ 
\begin{array}{c}
1\,,\ \ z>0\,, \\ 
1/2\,,\ \ z=0\,, \\ 
0\,,\ \ z<0\,.%
\end{array}%
\right. \,,\ \ \frac{d}{dz}\theta \left( z\right) =\delta \left( z\right) \,,
\end{equation*}%
and $\delta \left( z\right) $ stands for the Dirac delta function. Eq. (\ref%
{1.1}) supplies us with the simplest example, where the magnetic field
comprising the nonlinear vacuum respond can be explicitly studied, the shape
of the lines of force being fully described. If not for the linear
polarization, the potential (\ref{1.1}) would be the field of extended
spherically-symmetric charge \bigskip distributed with the constant density $%
\rho \left( r\right) $ inside a sphere $r\leq R$ with the radius $R$:%
\begin{equation}
\rho \left( r\right) =\left( \frac{3}{4\pi }\frac{Ze}{R^{3}}\right) \theta
\left( r-R\right) \,.
\end{equation}%
\ However, it should be kept in mind that with the account of the linear
vacuum polarization, the potential distribution (\ref{1.1}) cannot be
supported by any spherically-symmetric charge, strictly localized in a
finite space domain. To find the genuine source of the field (\ref{1.1}),
one should apply the operator in the left-hand side of (\ref{alin}) to it.
The result looks like:%
\begin{equation*}
\rho _{\text{lin}}\left( \mathbf{x}\right) =\rho \left( r\right) (1-\mathcal{%
L}_{\mathfrak{F}})+2\mathfrak{F}\mathcal{L}_{\mathfrak{GG}}\left( 1+\frac{(%
\mathbf{B\cdot x})^{2}}{B^{2}}\frac{d}{rdr}\right) \frac{d}{rdr}a_{0}\left(
r\right) .
\end{equation*}%
This charge density is cylindrically symmetric and extends beyond the
sphere, $r>R,$ decreasing as 1/$r^{3},$ or 1/$x_{3}^{3}$ far from it,
depending on the direction.

In the next Section the particular result to be formulated in the present
Section that the magnetic response to an electrostatic field implies that
the charge giving rise to the latter carries a magnetic dipole moment
parallel to the external magnetic field will be confirmed for a general
charge density distribution and for its, cylindrically-symmetric in the
remote domain, electrostatic field, where the linear response is also
included. Only the expression for the magnetic moment will be less explicit,
as well as expressions for the induced magnetic field and the shape of its
lines of force in the region closer to the charge. The aim of the present
Section is just to detail these appealing to the simplified example of (\ref%
{1.1}).

Now we proceed with the applied potential (\ref{1.1}). It provides the
following electric field 
\begin{eqnarray}
&&\mathbf{E}\left( \mathbf{x}\right) =\left(\frac{Ze}{4\pi}\right) \mathcal{E%
}\left( r\right) \mathbf{x}\,,  \label{E} \\
&&\mathcal{E}\left( r\right) =\frac{\theta \left( R-r\right) }{R^{3}}+\frac{%
\theta \left( r-R\right) }{r^{3}}\,.  \label{E1}
\end{eqnarray}%
In writing the expression for (\ref{E1}) we took into account the continuity
of (\ref{1.1}) and of its derivatives at $r=R$. \ As a result, Dirac-delta
terms stemming from the differentiation of the step function could be
omitted (see Appendix for a general discussion on the subject). This
simplification will be used every time functions are continuous.

\subsection{Nonlinearly induced current}

Taking into account the electric field (\ref{E}) and%
\begin{equation}
\mathcal{E}^{2}\left( r\right) =\frac{\theta \left( R-r\right) }{R^{6}}+%
\frac{\theta \left( r-R\right) }{r^{6}}\,,  \label{3.1}
\end{equation}
contributions (\ref{3})\ into the nonlinear current take the form,%
\begin{eqnarray}
&&j_{i}^{_{\mathfrak{FF}}}\left( \mathbf{x}\right) =\mathcal{L}_{\mathfrak{FF%
}}\varepsilon _{ijk}B_{k}E_{l}\partial _{j}E_{l}\,,  \notag \\
&&j_{i}^{_{\mathfrak{FGG}}}\left( \mathbf{x}\right) =-\mathcal{L}_{\mathfrak{%
FGG}}\varepsilon _{ijk}B_{k}B_{l}B_{n}E_{l}\left( \partial _{j}E_{n}\right)
\,,  \notag \\
&&j_{i}^{\mathfrak{GG}}\left( \mathbf{x}\right) =-\mathcal{L}_{\mathfrak{GG}%
}\varepsilon _{ijk}\partial _{j}\left( E_{k}B_{l}E_{l}\right) .  \label{4.1}
\end{eqnarray}
Therefore%
\begin{eqnarray}
\mathbf{j}_{\mathrm{nl}}\left( \mathbf{x}\right) &=&\left(\frac{Ze}{4\pi}%
\right) ^{2}\left\{ \left( \mathcal{L}_{\mathfrak{FF}}+\mathcal{L}_{%
\mathfrak{GG}}\right) \frac{\theta \left( R-r\right) }{R^{6}}\right.  \notag
\\
&+&\left. \left( \mathcal{L}_{\mathfrak{GG}}-2\mathcal{L}_{\mathfrak{FF}}+3%
\mathcal{L}_{\mathfrak{FGG}}\frac{\left( \mathbf{B}\cdot \mathbf{x}\right)
^{2}}{r^{2}}\right) \frac{\theta \left( r-R\right) }{r^{6}}\right\} \left( 
\mathbf{x\times B}\right) \,.  \label{4.2}
\end{eqnarray}%
Again, thanks to the continuity of $\mathbf{E}\left( \mathbf{x}\right) $ (%
\ref{E}), expressions (\ref{4.1}), (\ref{4.2}) do not contain the $\delta $%
-like contributions that might come from the differentiation of the
Heaviside $\theta $-function. But the nonlinear current is discontinuous at
the surface of the sphere $r=R$. We shall see below that the magnetic field
produced by it is still continuous.

The lines of the nonlinear current (\ref{4.2}) are circular and lie in
planes, orthogonal to the external magnetic field. Its density decreases as
the sixth power of the distance from the center of the charge, or of the
distance from the axis, parallel to the external magnetic field -- for large
distances.

Eq. (\ref{4.2}) is our final expression for the current that may be of
independent interest. In calculating the magnetic field produced by it in
the next Subsection, however, we shall not exploit it, but refer to
expressions of the previous Sections.

Observe that the proportionality of the current $\mathbf{j}_{\mathrm{nl}%
}\left( \mathbf{x}\right) $ above to $\left( \mathbf{x\times B}\right) ,$
and hence its circular character in the transverse plane, is valid also for
any spherically-symmetric field distribution like (\ref{E}), irrespective of
the special form (\ref{E1}). This property means that an expansion of (\ref%
{3}) over the (spacial part of) eigenvectors of the polarization operator in
a magnetic field (\ref{eigenvect}) does not contain, in the momentum space,
a nonvanishing contribution proportional to the vector $\mathbf{b}^{(1)},$
whose components are\textit{\ } $\flat _{1,2}^{(1)}=k_{1,2}k_{3}^{2},$ $%
\flat _{3}^{(1)}=-k_{3}(k_{1}^{2}+k_{2}^{2}),$ but is proportional to the
vector $\mathbf{b}^{(3)},$ such that $\flat _{3}^{(3)}=0,$ $\flat
_{1}^{(3)}=-k_{2},$ $\flat _{2}^{(3)}=k_{1},$ which is thereby the only
vector contributing in the expansion of the nonlinear current in
spherically-symmetric case. [The three vectors $\mathbf{b}^{(1)},\ \mathbf{b}%
^{(3)},$and $\mathbf{b}^{(4)}=\mathbf{k}$ are mutually orthogonal. There is
no contribution proportional to $\mathbf{b}^{(4)}$ due to the continuity $%
\mathbf{\nabla j}_{\text{\textrm{nl}}}=0.$ As for the eigenvector $\flat
_{\mu }^{(2)},$ its spacial part is zero in our static case of $k_{0}=0.$]
Correspondingly, only the value $c=3$ appears in the expansion of the
nonlinear magnetic field (\ref{1,3}).\textbf{\ }We do not know, if this
circumstance may be a general consequence of the spherical symmetry,
independent of the infrared approximation.

\subsection{Nonlinearly induced magnetic field}

We shall calculate here the magnetic field, nonlinearly induced by the
electrostatic field (\ref{1.1}) within the infrared approximation, basing on
Eq. (\ref{magfield}), which assumes the neglect of the linear vacuum
polarization.

\subsubsection{Second contribution in (\protect\ref{magfield})}

In order to find the integrals $\mathfrak{I}_{k}\left( \mathbf{x}\right) =%
\mathfrak{I}_{k}^{\mathfrak{FF}}\left( r\right) +\mathfrak{I}_{k}^{\mathfrak{%
FGG}}\left( \mathbf{x}\right) +\mathfrak{I}_{k}^{\mathfrak{GG}}\left( 
\mathbf{x}\right) $ it should be noted that the three their parts have the
following structure:%
\begin{eqnarray}
&&\mathfrak{I}_{k}^{\mathfrak{FF}}\left( r\right) =\int d^{3}y\frac{%
\mathfrak{h}_{k}^{\mathfrak{FF}}\left( y\right) }{\left\vert \mathbf{x}-%
\mathbf{y}\right\vert }=\frac{1}{2}\left(\frac{Ze}{4\pi}\right)^{2}\mathcal{L%
}_{\mathfrak{FF}}B_{k}v\left( r\right) \,,\,\, y=\left\vert\mathbf{y}%
\right\vert\,,  \notag \\
&&\mathfrak{I}_{k}^{\mathfrak{FGG}}\left( \mathbf{x}\right) =\int d^{3}y%
\frac{\mathfrak{h}_{k}^{\mathfrak{FGG}}\left( \mathbf{y}\right) }{\left\vert 
\mathbf{x}-\mathbf{y}\right\vert }=-\frac{1}{2}\left(\frac{Ze}{4\pi}%
\right)^{2}\mathcal{L}_{\mathfrak{FGG}}B_{k}\left[ (\mathbf{B\cdot x)}%
^{2}u(r)+B^{2}w(r)\right] \,,  \notag \\
&&\mathfrak{I}_{k}^{\mathfrak{GG}}\left( \mathbf{x}\right) =\int d^{3}y\frac{%
\mathfrak{h}_{k}^{\mathfrak{GG}}\left( \mathbf{y}\right) }{\left\vert 
\mathbf{x}-\mathbf{y}\right\vert }=-\left(\frac{Ze}{4\pi}\right)^{2}\mathcal{%
L}_{\mathfrak{GG}}\left[ x_{k}(\mathbf{B\cdot x)}u(r)+B_{k}w(r)\right] \,,
\label{5.3}
\end{eqnarray}%
where%
\begin{equation}
v\left( r\right) =\int d^{3}y\frac{\mathcal{E}^{2}\left( y\right) y^{2}}{%
\left\vert \mathbf{x}-\mathbf{y}\right\vert }\,,  \label{5.3a}
\end{equation}%
and $u(r)$ and $w(r)$ are the scalar coefficients in the tensor decomposition%
\begin{equation}
\int d^{3}y\frac{\mathcal{E}^{2}\left( y\right) y_{i}y_{k}}{\left\vert 
\mathbf{x}-\mathbf{y}\right\vert }=u\left( r\right) x_{i}x_{j}+w\left(
r\right) \delta _{ij}\,.  \label{5.3b}
\end{equation}

The basic angular integrals involved in (\ref{5.3a}), (\ref{5.3b}) are (we
refer to Eq. (2.222) in \cite{Gradshtein} for their values):

\begin{equation}
V_{1}(r,y)=\int_{-1}^{1}d\left( \cos \vartheta \right) \frac{1}{\sqrt{%
r^{2}+y^{2}-2ry\cos \vartheta }}=\frac{1}{ry}\left\{ r+y-\left\vert
r-y\right\vert \right\} \,,\text{ \ }r,y\geqslant 0\,,  \label{V.1}
\end{equation}%
and%
\begin{eqnarray}
V_{2}(r,y) &=&\int_{-1}^{1}d\left( \cos \vartheta \right) \frac{\cos
^{2}\vartheta }{\sqrt{r^{2}+y^{2}-2ry\cos \vartheta }}  \notag \\
&=&\left( \frac{2r^{4}-2r^{3}y+7r^{2}y^{2}-2ry^{3}+2y^{4}}{15r^{3}y^{3}}%
\right) (r+y)  \notag \\
&-&\left( \frac{2r^{4}+2r^{3}y+7r^{2}y^{2}+2ry^{3}+2y^{4}}{15r^{3}y^{3}}%
\right) \left\vert r-y\right\vert \,,\text{ \ }r,y\geqslant 0\,.  \label{V.2}
\end{eqnarray}%
One can immediately verify that $v\left( r\right) $ (\ref{5.3a}) takes the
form%
\begin{equation}
v\left( r\right) =2\pi \int_{0}^{\infty }dyy^{4}\mathcal{E}^{2}\left(
y\right) V_{1}(r,y)\,.  \label{v}
\end{equation}%
Further, once (\ref{5.3b}) does not depend on $\mathbf{B}$, one can choose $%
\mathbf{B\perp x}$ to find,%
\begin{equation}
w(r)=\frac{1}{\mathbf{B}^{2}}\int d^{3}y\left. \frac{\mathcal{E}^{2}\left(
y\right) (\mathbf{B\cdot y)}^{2}}{\left\vert \mathbf{x}-\mathbf{y}%
\right\vert }\right\vert _{_{\mathbf{B\perp x}}}=\pi \int_{0}^{\infty }dy%
\mathcal{E}^{2}\left( y\right) \left( V_{1}(r,y)-V_{2}(r,y)\right) \,.
\label{W}
\end{equation}

When getting this result we counted off the angle $\vartheta $ from the
radius-vector $\mathbf{x,}$ and $\varphi $ from $\mathbf{B}$, so that $%
\mathbf{B\cdot y}=By\sin \vartheta \cos \varphi $, and took into account the
relation $\int_{0}^{2\pi }\cos ^{2}\varphi d\varphi =\pi $. Analogously, by
choosing $\mathbf{B\parallel x}$ we find from (\ref{5.3b}) that%
\begin{eqnarray}
r^{2}u\left( r\right) +w\left( r\right) &=&\frac{1}{\mathbf{B}^{2}}\int
d^{3}y\left. \frac{\mathcal{E}^{2}\left( y\right) (\mathbf{B\cdot y)}^{2}}{%
\left\vert \mathbf{x}-\mathbf{y}\right\vert }\right\vert _{_{\mathbf{%
B\parallel x}}}  \label{v+w} \\
&=&2\pi \int_{0}^{\infty }dyy^{4}\mathcal{E}^{2}\left( y\right) V_{2}(r,y)\,,
\notag
\end{eqnarray}%
since now $\mathbf{B\cdot y=}By\cos \vartheta .$ Using Eq. (\ref{W}) for $%
w\left( r\right) $ we get the function $u\left( r\right) $ in (\ref{5.3b}),%
\begin{equation}
u(r)=\frac{\pi }{r^{2}}\int_{0}^{\infty }dyy^{4}\mathcal{E}^{2}\left(
y\right) \left( 3V_{2}(r,y)-V_{1}(r,y)\right) \,.  \label{U}
\end{equation}%
Now Eqs. (\ref{v}),\ (\ref{W}), and (\ref{U}) define the contributions (\ref%
{5.3}) into the field $\mathbf{h}\left( \mathbf{x}\right) $ according to (%
\ref{magfield}). These equations are valid, generally, for arbitrary
spherically symmetric field distribution of the form (\ref{E}), provided
that $\mathcal{E}\left( r\right) $ there decreases sufficiently fast at
large $r$ to guarantee the convergence of the remaining integrals over $y$.
For $\mathcal{E}^{2}\left( y\right) $ taken as (\ref{3.1}) the remaining $y$%
-integrations in (\ref{v}) and (\ref{v+w}) can be explicitly done with the
help of (\ref{V.1}), (\ref{V.2}), and their calculation is illustrated in
Appendix.

We obtain in this way 
\begin{equation}
v\left( r\right) =\pi \left[ \frac{3}{R^{2}}\left( 1-\frac{r^{4}}{15R^{4}}%
\right) \theta \left( R-r\right) +\frac{2}{r^{2}}\left( \frac{12r}{5R}%
-1\right) \theta \left( r-R\right) \right] \,,  \label{6a}
\end{equation}%
for (\ref{v}), and%
\begin{eqnarray}
r^{2}u\left( r\right) +w\left( r\right) &=&\frac{\pi }{R^{2}}\left( 1+\frac{%
2r^{2}}{5R^{2}}-\frac{9r^{4}}{35R^{4}}\right) \theta \left( R-r\right) 
\notag \\
&+&\frac{8\pi }{5Rr}\left( 1-\frac{2R^{2}}{7r^{2}}\right) \theta \left(
r-R\right) \,,  \notag \\
u\left( r\right) &=&\frac{3\pi }{5R^{2}}\left( 1-\frac{10r^{2}}{21R^{2}}%
\right) \frac{\theta \left( R-r\right) }{R^{2}}+\pi \left( 1-\frac{24R}{35r}%
\right) \frac{\theta \left( r-R\right) }{r^{4}}\,,  \notag \\
w\left( r\right) &=&\pi \left( 1-\frac{r^{2}}{5R^{2}}+\frac{r^{4}}{35R^{4}}%
\right) \frac{\theta \left( R-r\right) }{R^{2}}  \notag \\
&-&\pi \left( 1-\frac{8r}{5R}-\frac{8R}{35r}\right) \frac{\theta \left(
r-R\right) }{r^{2}}  \label{6b}
\end{eqnarray}%
for (\ref{v+w}), (\ref{U}) and (\ref{W}). With these results the integrals (%
\ref{5.3}) have the form%
\begin{eqnarray}
&&\mathfrak{I}_{k}^{\mathfrak{FF}}\left( r\right) =\pi \left(\frac{Ze}{4\pi}%
\right)^{2}\mathcal{L}_{\mathfrak{FF}}B_{k}\left[ \mathcal{I}_{1}\left(
r\right) \theta \left( R-r\right) +\mathcal{I}_{2}\left( r\right) \theta
\left( r-R\right) \right] \,,  \label{12.1} \\
&&\mathfrak{I}_{k}^{\mathfrak{FGG}}\left( \mathbf{x}\right) =\pi \left(\frac{%
Ze}{4\pi}\right)^{2}\mathcal{L}_{\mathfrak{FGG}}B_{k}\left[ \mathcal{I}%
_{3}\left( r\right) \theta \left( R-r\right) +\mathcal{I}_{4}\left( r\right)
\theta \left( r-R\right) \right] \,,  \label{12.2} \\
&&\mathfrak{I}_{k}^{\mathfrak{GG}}\left( \mathbf{r}\right) =\pi \left(\frac{%
Ze}{4\pi}\right)^{2}\mathcal{L}_{\mathfrak{GG}}\left[ \mathcal{I}%
_{k}^{5}\left( r\right) \theta \left( R-r\right) +\mathcal{I}_{k}^{6}\left(
r\right) \theta \left( r-R\right) \right] \,,  \label{12.3}
\end{eqnarray}%
where%
\begin{eqnarray}
\mathcal{I}_{1}\left( r\right) &=&\frac{3}{2R^{2}}\left( 1-\frac{r^{4}}{%
15R^{4}}\right) \,,\ \ \mathcal{I}_{2}\left( r\right) =-\frac{1}{r^{2}}%
\left( 1-\frac{12r}{5R}\right) \,,  \notag \\
\mathcal{I}_{3}\left( r\right) &=&-\frac{1}{2R^{2}}\left[ \left( 1-\frac{%
r^{2}}{5R^{2}}+\frac{r^{4}}{35R^{4}}\right) B^{2}+\frac{3r^{2}}{5R^{2}}%
\left( 1-\frac{10r^{2}}{21R^{2}}\right) \left( \frac{\mathbf{B\cdot x}}{r}%
\right) ^{2}\right] \,,  \notag \\
\mathcal{I}_{4}\left( r\right) &=&-\frac{4}{5Rr}\left[ \left( 1+\frac{R^{2}}{%
7r^{2}}-\frac{5R}{8r}\right) B^{2}+\frac{5R}{8r}\left( 1-\frac{24R}{35r}%
\right) \left( \frac{\mathbf{B\cdot x}}{r}\right) ^{2}\right] \,,  \notag \\
\mathcal{I}_{k}^{5}\left( r\right) &=&-\frac{3x_{k}(\mathbf{B\cdot x)}}{%
5R^{4}}\left( 1-\frac{10r^{2}}{21R^{2}}\right) -\frac{B_{k}}{R^{2}}\left( 1-%
\frac{r^{2}}{5R^{2}}+\frac{r^{4}}{35R^{4}}\right) \,,  \notag \\
\mathcal{I}_{k}^{6}\left( r\right) &=&-\frac{x_{k}(\mathbf{B\cdot x)}}{r^{4}}%
\left( 1-\frac{24R}{35r}\right) +\frac{B_{k}}{r^{2}}\left( 1-\frac{8r}{5R}-%
\frac{8R}{35r}\right) \,.  \label{13a-1}
\end{eqnarray}

Using these representations it is straightforward to make sure that the
functions (\ref{12.1}) -- (\ref{12.3}) are continuous in the point $r=R$. So
are also all their first and second derivatives with respect to the
coordinate components. Consequently, the Dirac delta-functions and their
derivatives, which stem from differentiation of the step-functions in the
calculation of $\partial _{i}\partial _{k}\mathfrak{I}_{k}\left( \mathbf{x}%
\right) /4\pi $ from (\ref{5.3}), do not contribute (see Appendix). As a
result these derivatives can be left in their simplest form,

\begin{eqnarray}
\frac{\partial _{i}\partial _{k}}{4\pi }\mathfrak{I}_{k}\left( \mathbf{x}%
\right) &=&\left(\frac{Ze}{8\pi}\right)^{2}\left\{ \partial_i \partial_k %
\left[ \mathcal{L}_{\mathfrak{FF}}B_{k}\mathcal{I}_{1}\left( r\right)
\right. +\left. \mathcal{L}_{\mathfrak{FGG}}B_{k} \mathcal{I}_{3}\left(
r\right) +\mathcal{L}_{\mathfrak{GG}}\mathcal{I}_{k}^{5}\left( r\right) %
\right] \theta \left( R-r\right) \right.  \notag \\
&+&\left. \partial_i \partial_k \left[ \mathcal{L}_{\mathfrak{FF}}B_{k}%
\mathcal{I}_{2}\left( r\right) +\mathcal{L}_{\mathfrak{FGG}}B_{k}\mathcal{I}%
_{4}\left( r\right) \right. + \left. \mathcal{L}_{\mathfrak{GG}}\mathcal{I}%
_{k}^{6}\left( r\right) \right] \right\} \theta \left( r-R\right) \,.
\label{13a-8}
\end{eqnarray}%
The final form of (\ref{13a-8}) is:%
\begin{eqnarray}
\frac{\partial _{i}\partial _{k}}{4\pi }\mathfrak{I}_{k}\left( \mathbf{x}%
\right) &=&\left(\frac{Ze}{8\pi}\right)^{2}\left\{ \frac{\theta \left(
R-r\right) }{R^{4}}\left[ 2\left( -\mathcal{L}_{\mathfrak{GG}}-\frac{1}{5}%
B^{2}\mathcal{L}_{\mathfrak{FGG}}\right) B_{i} \right. \right.  \notag \\
&-&\left. \frac{2r^{2}}{5R^{2}}\left( \left( B_{i}+\frac{2\left( \mathbf{B}%
\cdot \mathbf{x}\right) x_{i}}{r^{2}}\right) \left( \mathcal{L}_{\mathfrak{FF%
}}-4\mathcal{L}_{\mathfrak{GG}}-\frac{4}{7}B^{2}\mathcal{L}_{\mathfrak{FGG}%
}\right)-\frac{15}{7}\mathcal{L}_{\mathfrak{FGG}}\left( \frac{\mathbf{B\cdot
x}}{r}\right) ^{2}B_{i}\right) \right]  \notag \\
&+&\frac{\theta \left( r-R\right) }{r^{4}}\left[ 2\left( B_{i}-4\frac{\left( 
\mathbf{B}\cdot \mathbf{x}\right) x_{i}}{r^{2}}\right) \left( \mathcal{L}_{%
\mathfrak{FF}}-\mathcal{L}_{\mathfrak{GG}}-\mathcal{L}_{\mathfrak{FGG}%
}B^{2}\right) \right.  \notag \\
&+&6\mathcal{L}_{\mathfrak{FGG}}\left( \frac{\mathbf{B\cdot x}}{r}\right)
^{2}\left( B_{i}-2\frac{\left( \mathbf{B\cdot x}\right) x_{i}}{r^{2}}\right)
\notag \\
&-&\frac{4r}{5R}\left( B_{i}-3\frac{\left( \mathbf{B}\cdot \mathbf{x}\right)
x_{i}}{r^{2}}\right) \left( 3\mathcal{L}_{\mathfrak{FF}}-\mathcal{L}_{%
\mathfrak{FGG}}B^{2}-2\mathcal{L}_{\mathfrak{GG}}\right) +\frac{36R}{35r}%
\left( B_{i}-5\frac{\left( \mathbf{B\cdot x}\right) x_{i}}{r^{2}}\right)
B^{2}\mathcal{L}_{\mathfrak{FGG}}  \notag \\
&-&\left. \left. \frac{12R}{7r}\left( \frac{\mathbf{B\cdot x}}{r}\right)
^{2}\left( 3B_{i}-7\frac{\left( \mathbf{B\cdot x}\right) x_{i}}{r^{2}}%
\right) \mathcal{L}_{\mathfrak{FGG}}\right] \right\} \,.  \label{18b}
\end{eqnarray}

\subsubsection{Total nonlinear magnetic field}

With the explicit form (\ref{18b}) of $\partial _{i}\partial _{k}\mathfrak{I}%
_{k}\left( \mathbf{x}\right) /4\pi $ we proceed to evaluate the total
magnetic field (\ref{magfield}). Bearing in mind (\ref{E}) and (\ref{3.1})
the field $\mathfrak{h}_{i}\left( \mathbf{x}\right) $ in (\ref{hgerman}) is
written as%
\begin{eqnarray}
\mathfrak{h}_{i}\left( \mathbf{x}\right) &=&\left( \frac{Ze}{8\pi}\right)
^{2}\left\{ \frac{2r^{2}}{R^{2}}\left[ \left( \mathcal{L}_{\mathfrak{FF}}-%
\mathcal{L}_{\mathfrak{FGG}}\left( \frac{\mathbf{B\cdot x}}{r}\right)
^{2}\right) B_{i} - 2\mathcal{L}_{\mathfrak{GG}}\frac{\left( \mathbf{B}\cdot 
\mathbf{x}\right) x_{i}}{x^{2}}\right] \frac{\theta \left( R-r\right) }{R^{4}%
} \right.  \notag \\
&+&\left. 2\left[ \left( \mathcal{L}_{\mathfrak{FF}}-\mathcal{L}_{\mathfrak{%
FGG}}\left( \frac{\mathbf{B\cdot x}}{r}\right) ^{2}\right) B_{i}-2\mathcal{L}%
_{\mathfrak{GG}}\frac{\left( \mathbf{B}\cdot \mathbf{x}\right) x_{i}}{r^{2}}%
\right] \frac{\theta \left( r-R\right) }{r^{4}}\right\} \,,  \label{19a}
\end{eqnarray}%
and the total magnetic field (\ref{magfield}) has the final form%
\begin{eqnarray}
h_{i}\left( \mathbf{x}\right) &=&h_{i}^{\mathrm{in}}\left( \mathbf{x}\right)
\theta \left( R-r\right) +h_{i}^{\mathrm{out}}\left( \mathbf{x}\right)
\,\theta \left( r-R\right) ,  \label{19b} \\
h_{i}^{\mathrm{in}}\left( \mathbf{x}\right) &=&h_{i}^{\mathrm{in}}\left( 
\mathbf{x}\right) _{\perp }-\frac{2r^{2}}{R^{2}}\left( \frac{Ze}{4\pi R^{2}}%
\right) ^{2}\left\{ \frac{1}{7}\mathcal{L}_{\mathfrak{FGG}}\left( \frac{%
\mathbf{B\cdot x}}{r}\right) ^{2}B_{i}\right.  \notag \\
&+&\left. \frac{1}{5}\left( \frac{1}{2}\mathcal{L}_{\mathfrak{FF}}+\frac{1}{2%
}\mathcal{L}_{\mathfrak{GG}}-\frac{2}{7}B^{2}\mathcal{L}_{\mathfrak{FGG}%
}\right) \frac{\left( \mathbf{B}\cdot \mathbf{x}\right) x_{i}}{r^{2}}%
\right\} \,,  \label{18.7} \\
h_{i}^{\mathrm{in}}\left( \mathbf{x}\right) _{\perp } &=&-\left( \frac{Ze}{%
4\pi R^{2}}\right) ^{2}\left[ \frac{1}{2}\left( 1-\frac{4r^{2}}{5R^{2}}%
\right) \mathcal{L}_{\mathfrak{GG}}+\frac{x^{2}}{10R^{2}}\mathcal{L}_{%
\mathfrak{FF}}+ \frac{1}{10}\left( 1-\frac{4r^{2}}{7R^{2}}\right) B^{2}%
\mathcal{L}_{\mathfrak{FGG}}\right] B_{i}\,,  \label{perp.1} \\
h_{i}^{\mathrm{out}}\left( \mathbf{x}\right) &=&h_{i}^{\mathrm{out}}\left( 
\mathbf{x}\right) _{\perp }+\left( \frac{Ze}{4\pi r^{2}}\right) ^{2}\left( 1-%
\frac{9R}{7r}\right) \mathcal{L}_{\mathfrak{FGG}}\left( \frac{\mathbf{B\cdot
x}}{r}\right) ^{2}B_{i}  \notag \\
&-&2\left( \frac{Ze}{4\pi r^{2}}\right) ^{2}\left\{ \left( 1-\frac{9r}{10R}%
\right) \mathcal{L}_{\mathfrak{FF}}-\frac{1}{2}\left( 1-\frac{6r}{5R}\right) 
\mathcal{L}_{\mathfrak{GG}}\right.  \notag \\
&+&\left. \left[ \left( -1+\frac{3r}{10R}+\frac{9R}{14r}\right) B^{2}+\frac{3%
}{2}\left( 1-\frac{R}{r}\right) \left( \frac{\mathbf{B\cdot x}}{r}\right)
^{2}\right] \mathcal{L}_{\mathfrak{FGG}}\right\} \frac{\left( \mathbf{B}%
\cdot \mathbf{x}\right) x_{i}}{r^{2}}\,,  \label{18.6} \\
h_{i}^{\mathrm{out}}\left( \mathbf{x}\right) _{\perp } &=&\left( \frac{Ze}{%
4\pi r^{2}}\right) ^{2}\left\{ \frac{1}{2}\left( 1-\frac{6r}{5R}\right) 
\mathcal{L}_{\mathfrak{FF}}-\frac{1}{2}\left( 1-\frac{4r}{5R}\right) 
\mathcal{L}_{\mathfrak{GG}}- \frac{1}{2}\left( 1-\frac{2r}{5R}-\frac{18R}{35r%
}\right) \mathcal{L}_{\mathfrak{FGG}}B^{2}\right\} B_{i}\,.  \notag
\end{eqnarray}

Here $h_{i}^{\mathrm{in}}\left( \mathbf{x}\right) $ represents the total
magnetic field for points inside the sphere $\left( r<R\right) $ while the
designation $h_{i}^{\mathrm{out}}\left( \mathbf{x}\right) $ is reserved to
the field outside the sphere $\left( r>R\right) $. The total magnetic field (%
\ref{19b}) is continuous at $r=R$. The orientation of the magnetic field (%
\ref{19b}) will be revealed in the next Subsection, where we present the
shape of the lines of magnetic field.

The long-range contribution of (\ref{18.6}), $h_{i}^{\mathrm{LR}}\left( 
\mathbf{x}\right) $, behaves like a magnetic field generated by a magnetic
dipole:%
\begin{equation}
h_{i}^{\mathrm{LR}}\left( \mathbf{x}\right) =\frac{3\left( \mathbf{x}\cdot 
\boldsymbol{\mu }\right) x_{i}}{r^{5}}-\frac{\mu _{i}}{r^{3}}\,,
\label{18.9}
\end{equation}%
with $\boldsymbol{\mu }$ being the equivalent magnetic dipole moment, given
by%
\begin{equation}
\mu _{i}=\left(\frac{Ze}{4\pi}\right)^2\frac{1}{5R}\left( 3\mathcal{L}_{%
\mathfrak{FF}}-2\mathcal{L}_{\mathfrak{GG}}-B^{2}\mathcal{L}_{\mathfrak{FGG}%
}\right) B_{i}\,.  \label{18.10}
\end{equation}

\subsection{Magnetic moment of a spherical charge}

In this section we are going to explore the dependence of the magnetic
moment (\ref{18.10}) with respect to the external applied magnetic field $B$%
. To do that one has to consider the corresponding dependence of the
coefficients $\mathcal{L}_{\mathfrak{FF}},\,\mathcal{L}_{\mathfrak{GG}}$ and 
$\mathcal{L}_{\mathfrak{FGG}}$ on $B$. Such coefficients essentially depend
on the model in consideration but, confining ourselves to QED and working within the local limit approximation, they have a specific form provided by the
Euler-Heisenberg effective Lagrangian \cite{Heisenberg-Euler}. They have
been considered before in \cite{Shabus2011} and due to this fact we use here
the expressions previously derived there to obtain\footnote{%
See equations (62) and (63) in \cite{Shabus2011}.},

\begin{equation}
3\mathcal{L}_{\mathfrak{FF}}-2\mathcal{L}_{\mathfrak{GG}}-B^{2}\mathcal{L}_{%
\mathfrak{FGG}}=\left( \frac{\alpha }{\pi B_{\mathrm{Sch}}^{2}b^{3}}\right)
\int_{0}^{\infty }dte^{-\frac{t}{b}}\left\{ \coth ^{2}t-\frac{(t^{2}+3)\coth t}{3t}\right\}<0 \,, \label{new4.2}
\end{equation}%
where $B_{\mathrm{Sch}}=m^2/e$ and $b=B/B_{\mathrm{Sch}}$. Using the latter result, the magnetic moment (\ref{18.10}) has the form,

\begin{equation}
\mu=\frac{\lambda}{b^2}\int_{0}^{\infty}e^{-\frac{t}{b}}\left(\coth^2 t
- \frac{(t^2 +3)\coth t}{3t}\right)\,,\,\,\,\lambda\equiv\left(\frac{Ze}{%
4\pi}\right)^2\left(\frac{\alpha}{5\pi R B_{\mathrm{Sch}}}\right)\,.  \label{new5}
\end{equation}
The negativity of (\ref{new4.2}) and of $\mu$ (\ref{new5}) indicates that the magnetic moment is directed opposite to the background magnetic field.

This integral does not have an analytical solution. To show the dependence
of the magnetic moment with respect to the external field we plot the
numerical results of the ratio $-\mu/\lambda$ for each given value
of $b$ within the range $10^{-2}\leq b\leq 50$ (Fig. \ref%
{magnmoment}). Although (\ref{new5}) does not have an analytical solution one, can estimate its asymptotic behaviours for small and large values of the external magnetic field $B$. In the first case, for $t$
sufficiently small, the exponent $e^{-t/b}$ is approximately zero. Separating the integral above in two parts, where the limit $\epsilon$ is chosen such as $b\ll \epsilon$, one can write 
\begin{align}
\frac{b^2}{\lambda}\mu&=\int_{0}^{\epsilon}e^{-\frac{t}{b}%
}f(t)+\int_{\epsilon}^{\infty}e^{-\frac{t}{b}}f(t)  \notag \\
&\simeq \int_{0}^{\epsilon}e^{-\frac{t}{b}}f(t)\,,\,\,\,\,\,f(t)=\coth^2 t - 
\frac{(t^2 +3)\coth t}{3t}\,,  \label{new6}
\end{align}
since the second integral in the first line is pratically zero ($t$ is
always $t\gg b$). Once the function $f(t)$ has
a maximum at $t=0$, one can conclude that the greatest contribution for (\ref{new6}) is
\begin{equation}
\frac{\mu}{\lambda}\simeq \frac{1}{b^2}\int_{0}^{\epsilon}e^{-\frac{t}{b}%
}\left(-\frac{t^2}{45}-\frac{t^4}{189}\right)\simeq \int_{0}^{\epsilon}e^{-%
\frac{t}{b}}\left(-\frac{t^2}{45}\right)\,,  \label{new7}
\end{equation}
which, after two integration by parts, we obtain 
\begin{equation}
\int_{0}^{\epsilon}e^{-\frac{t}{b}}\left(-\frac{t^2}{45}\right)\simeq \frac{1%
}{45}\left[2b^{3}-e^{-\frac{\epsilon}{b}}\left(b\epsilon^{2}+2b^2%
\epsilon+2b^3\right)\right]\,.  \label{new8}
\end{equation}
Finally the asymptotic form of (\ref{new5}) is linear in $b$, 
\begin{equation}
\mu\simeq -\lambda\left(\frac{2b}{45}\right)\,. \label{78}
\end{equation}

In the large-field asymptotic regime $B>>B_{Sch}$ integral (68)decreases as $ -\frac{\alpha}{3\pi}\frac{e}{m^2B}$ providing in turn a constant value to the magnetic moment (see the horizontal dot-dashed line in Fig.1) $\mu=-\lambda/
3$.

\begin{figure}[th!]
\begin{center}
\includegraphics[scale=0.48]{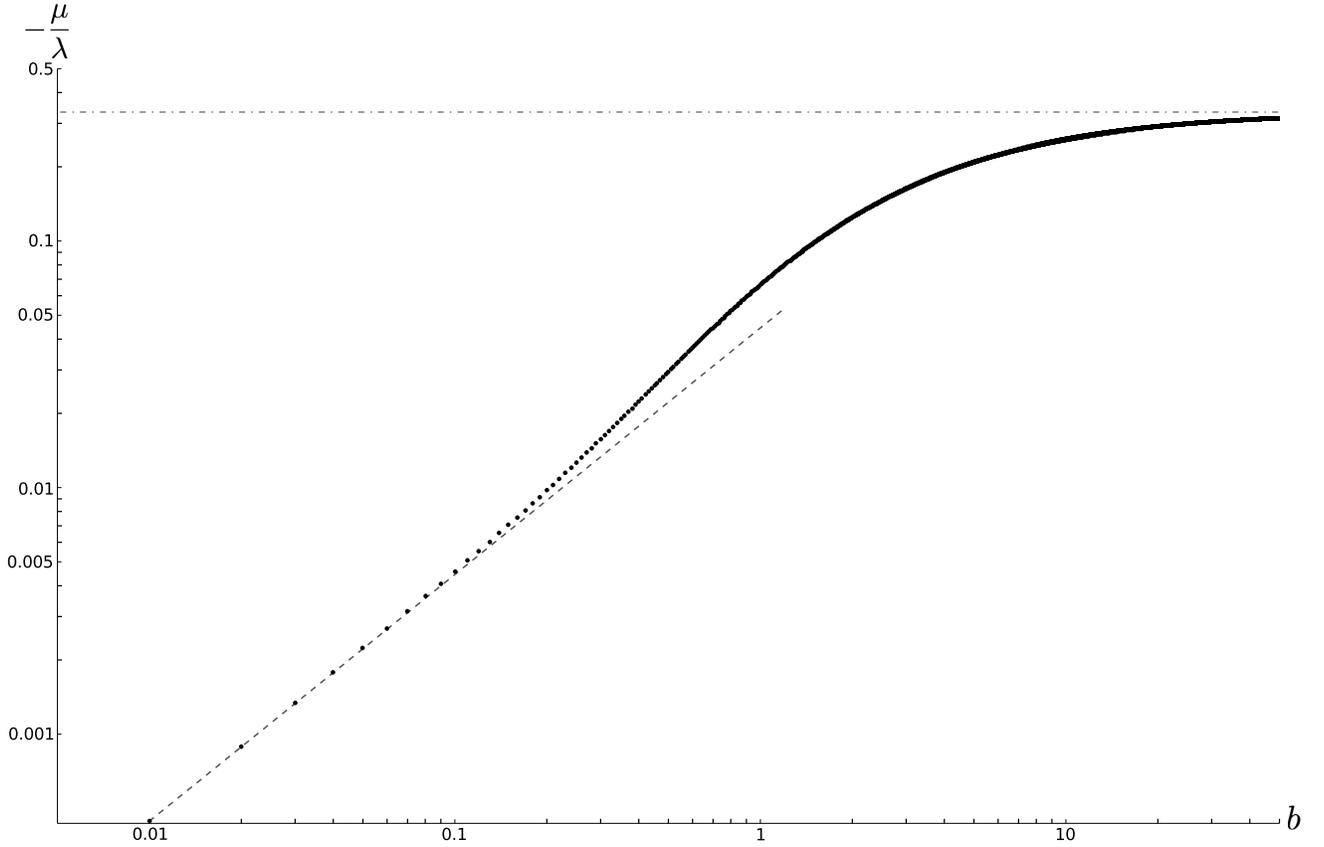}
\end{center}
\caption{The magnetic moment (\ref{new5}) of a charge plotted in logarithmic scale against the magnetic field $b=B/B_{\mathrm{Sch}}$ in the range $10^{-2}<b<50$. The scaling parameter is $\lambda= (Ze/4\pi)^{2}(\alpha/5\pi R B_{\mathrm{Sch}})$. The dot-dashed line corresponds to the large-field constant asymptotic value, while the dashed line is the small-field linear asymptotic behavior (\ref{78}).  The step in $b$ is $10^{-2}$. The leftmost value for the magnetic moment is approximately $-4.4 \times 10^{-3}\lambda$.}
\label{magnmoment}
\end{figure}

\subsection{Magnetic lines of force}

\subsubsection{Interior}

Here we are going to establish the form of the lines of force of (\ref{18.7}%
), first on the inside of the sphere. To this end, the same as in Sec. 2.1,
we direct axis $x_{3}$ along the magnetic field $\mathbf{B}$ and represent
the vector (\ref{18.7}) in the orthogonal basis of unit vectors $e_{\left(
1\right) i},$ $e_{\left( 3\right) i}$ directed along orthogonal axes 1 and 3
as%
\begin{equation}
h_{i}^{\mathrm{in}}\left( \mathbf{x}\right) =e_{\left( 3\right) i}B\left[ A+D%
\frac{r^{2}}{R^{2}}+\left( g+C\right) \left( \frac{x_{3}}{R}\right) ^{2}%
\right] +e_{\left( 1\right) i}BC\frac{x_{3}x_{1}}{R^{2}},  \label{basis}
\end{equation}
with 
\begin{eqnarray}
&&A=-\frac{1}{2}\left( \frac{Ze}{4\pi R^{2}}\right) ^{2}\left( \mathcal{L}_{%
\mathfrak{GG}}+\frac{1}{5}B^{2}\mathcal{L}_{\mathfrak{FGG}}\right) \,, 
\notag \\
&&D=\frac{2}{5}\left( \frac{Ze}{4\pi R^{2}}\right) ^{2}\left( \mathcal{L}_{%
\mathfrak{GG}}-\frac{1}{4}\mathcal{L}_{\mathfrak{FF}}+\frac{1}{7}B^{2}%
\mathcal{L}_{\mathfrak{FGG}}\right) \,,  \notag \\
&&g=-\frac{2}{7}\left( \frac{Ze}{4\pi R^{2}}\right) ^{2}\mathcal{L}_{%
\mathfrak{FGG}}B^{2}\,,  \label{ADg} \\
&&C=-\frac{1}{5}\left( \frac{Ze}{4\pi R^{2}}\right) ^{2}\left( \mathcal{L}_{%
\mathfrak{FF}}+\mathcal{L}_{\mathfrak{GG}}-\frac{4}{7}B^{2}\mathcal{L}_{%
\mathfrak{FGG}}\right) \,,  \label{C}
\end{eqnarray}%
being functions of $B$, independent of the coordinates $\mathbf{x}$. We set $%
x_{2}=0$, since the full pattern of the lines of force is to be obtained
from the one in the plane (3,1) by rotating along axis 3 due to cylindric
symmetry of the problem, so $r^{2}=x_{1}^{2}+x_{3}^{2}.$ Equalizing the
derivative $\frac{dx_{3}}{dx_{1}}$ with the ratio $\frac{h_{3}^{\mathrm{in}}%
}{h_{1}^{\mathrm{in}}}$ supplies us with the differential equation for the
shape of the line of force $x_{3}(x_{1}).$ With the new notations $y=x_{3}/R$
and $z=x_{1}/R$ this differential equation follows from (\ref{basis}) to be%
\begin{eqnarray}
y\frac{dy}{dz} &=&\frac{\beta +\gamma z^{2}+Ey^{2}}{z}\,,  \label{Solut} \\
\text{where }\beta &=&\frac{A}{C}\,,\ \ \gamma =\frac{D}{C}\,,\ \ E=\frac{D+g%
}{C}+1\,.  \notag
\end{eqnarray}%
This is the so-called second-type Abel first-order differential equation
with the family of solutions \cite{Kamke}

\begin{equation}
y=\sqrt{-\frac{\beta }{E}+\frac{\gamma z^{2}}{1-E}+\left( -\frac{z^{2}}{%
z_{0}^{2}}\right) ^{E}}\,,  \label{y}
\end{equation}%
parametrized by the integration constant $z_{0}$.

Extreme points of the lines of force given by (\ref{y}) are achieved at 
\newline
$z_{\text{\textrm{extr}}}^{2}=\left( \frac{\gamma }{\left( E-1\right) E}%
\right) ^{\frac{1}{E-1}}\left( -z_{0}^{2}\right) ^{\frac{E}{E-1}}.$ The
corresponding extremum value $y_{\text{extr}}$ of the vertical coordinate
turns to zero for the curve corresponding to the largest admitted value of
the integration constant, ($z_{0}^{2\text{ }})^{\text{foc}}=\gamma
^{-1}\beta ^{\frac{E-1}{E}}(E\left( 1-E\right) )^{\frac{1}{E}}.$ This closed
curve degenerates to a point. Its position at the abscissa axis is $z=z^{%
\text{\textrm{foc}}}=(\frac{-\beta }{\gamma })^{1/2}.$ We call this point
the focus of the lines-of-force pattern. Larger values of the integration
constant would not give rise to any line of force, since they would make $y$
complex for any $z.$ Therefore the integration constant can be taken within
the range $z_{0}^{\text{foc}}>z_{0}>0.$ As we let the parameter $z_{0}$
diminish down to the zero value, we pass to lines of force that go farther
and farther from the horizontal axis. In the limit $z_{0}=0$ we reach the
ultimate curve that passes through the origin $z=y=0$ and coincides with the
y-axis. The focal point may lie both inside and outside the sphere depending
on whether $\frac{-\beta }{\gamma }=\frac{-A}{D}$ is smaller or larger than
unity.

Bearing in mind the asymptotic behavior at large magnetic field $B\gg
m^{2}/e $

\begin{equation}
\mathcal{L}_{\mathfrak{FF}}=\frac{\alpha }{3\pi }\frac{1}{B^{2}}\,,\text{\ \ 
}\mathcal{L}_{\mathfrak{GG}}=\frac{\alpha }{3\pi }\left( \frac{e}{m^{2}}%
\right) \frac{1}{B}\,,\text{\ \ }B^{2}\mathcal{L}_{\mathfrak{FGG}}=B^{2}%
\frac{d\mathcal{L}_{\mathfrak{GG}}}{d\mathfrak{F}}=-\mathcal{L}_{\mathfrak{GG%
}}\,,  \label{numfactors}
\end{equation}%
of the basic quantities forming the coefficients $A,$ $D$ and $g$ (\ref{ADg}%
) in QED (see \textit{e.g.} \cite{Shabus2011}), we find for $z^{\text{%
\textrm{foc}}}$ the value $(7/6)^{1/2}>1$ outside the sphere in this limit.

In the limit of pure point-like dipole $\beta \rightarrow 0,$ the focal
point tends to the origin, and all the lines of force are squeezed between
these two points. In the large-field regime (\ref{numfactors}), the
constants are:%
\begin{eqnarray*}
\beta &=&\frac{2\mathcal{L}_{\mathfrak{GG}}}{\mathcal{L}_{\mathfrak{FF}}+%
\frac{11}{7}\mathcal{L}_{\mathfrak{GG}}}=\frac{2\left( \frac{e}{m^{2}}%
\right) }{\frac{1}{B}+\frac{11}{7}\left( \frac{e}{m^{2}}\right) }\,, \\
\gamma &=&\frac{\frac{1}{2}\mathcal{L}_{\mathfrak{FF}}-\frac{12}{7}\mathcal{L%
}_{\mathfrak{GG}}}{\mathcal{L}_{\mathfrak{FF}}+\frac{11}{7}\mathcal{L}_{%
\mathfrak{GG}}}=\frac{\frac{1}{2B}-\frac{12}{7}\left( \frac{e}{m^{2}}\right) 
}{\frac{1}{B}+\frac{11}{7}\left( \frac{e}{m^{2}}\right) }\,, \\
E &=&\frac{\frac{3}{2}\mathcal{L}_{\mathfrak{FF}}-\frac{11}{7}\mathcal{L}_{%
\mathfrak{GG}}}{\mathcal{L}_{\mathfrak{FF}}+\frac{11}{7}\mathcal{L}_{%
\mathfrak{GG}}}=\frac{\frac{3}{2B}-\frac{11}{7}\left( \frac{e}{m^{2}}\right) 
}{\frac{1}{B}+\frac{11}{7}\left( \frac{e}{m^{2}}\right) }\,.
\end{eqnarray*}

\begin{figure}[th!]
\begin{center}
\includegraphics[scale=0.6]{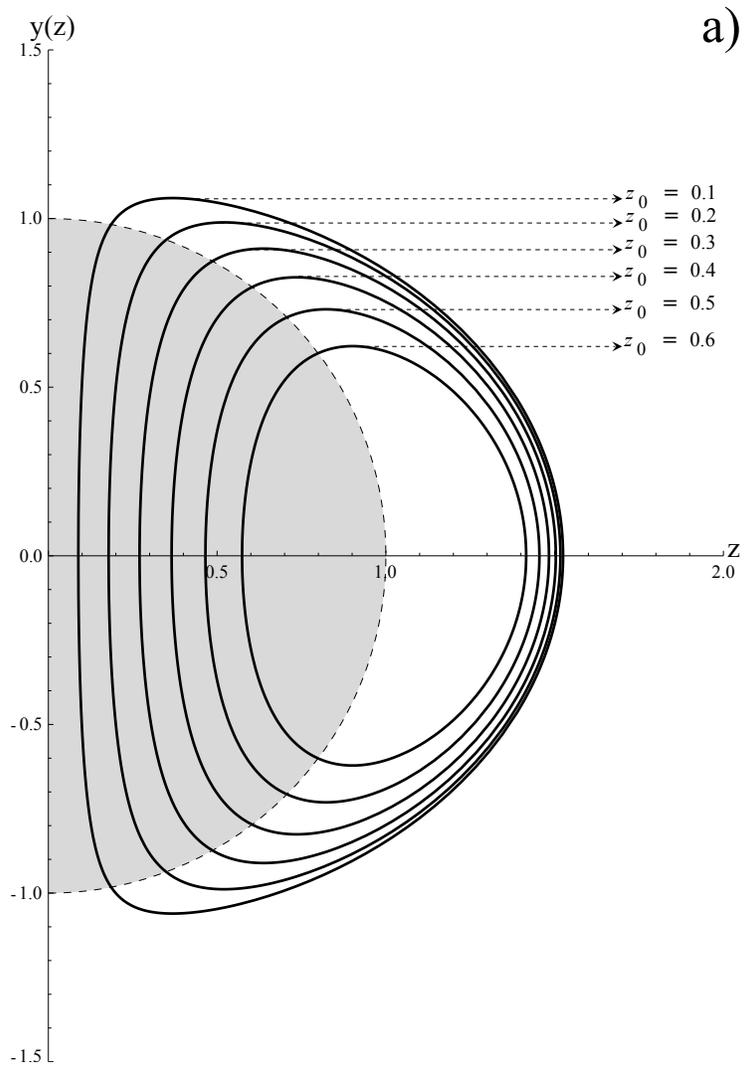}
\end{center}
\caption{Magnetic dipole lines of a static charge in external magnetic field
exampled with $B=\infty $. The pattern to be trusted inside the charge, $%
y^{2}+z^{2}<1$, following solution (\protect\ref{patinside}) with real $%
z_{0} $.}
\label{Figinside}
\end{figure}

For $B\rightarrow \infty $, the coefficients above become%
\begin{equation}
\beta =\frac{14}{11}\,,\ \ \gamma =-\frac{12}{11}\,,\ \ E=-1\,,
\label{const1}
\end{equation}%
and the magnetic curves take the final form%
\begin{equation}
y\left( z\right) =\sqrt{\frac{14}{11}-\frac{6}{11}z^{2}-\left( \frac{z_{0}}{z%
}\right) ^{2}}\,.  \label{patinside}
\end{equation}%
The family of magnetic lines labelled by positive values of the integration
constant $z_{0}$ in the interval $0<$ $z_{0}<\frac{7}{\sqrt{66}}$ is drawn
following Eq. (\ref{patinside}) in Fig. \ref{Figinside}. For negative $%
z_{0}^{2}$ the corresponding curves lie completely outside the sphere $%
z^{2}+y^{2}=1,$ rounding from outside the family presented in this Figure.
We are not interested in showing them, because our starting equations in
this Subsection belong to the interior of that sphere. For $z_{0}^{2}>\frac{%
49}{66}$ the solutions (\ref{patinside}) are no longer real. The values
taken for parametrizing the six curves in Fig. \ref{Figinside} are indicated
in the drawing. We must mistrust those parts of the curves in \ Fig. \ref%
{Figinside} that belong to the exterior of the sphere, and our next task is
to obtain the continuation of the magnetic lines of force to that region. 
\newpage

\subsubsection{Exterior}

Referring to the same basis and reference frame as in the previous
Subsection, the magnetic field outside the sphere (\ref{18.6}) reads 
\begin{eqnarray}
h_{i}^{\mathrm{out}}\left( \mathbf{x}\right) &=&B\left\{ \left[ \mathcal{A}%
^{\prime }+\left( \mathcal{B}^{\prime }+\mathcal{C}^{\prime }\right) y^{2}+%
\mathcal{D}^{\prime }y^{4}\right] e_{\left( 3\right) i}+\left( \mathcal{C}%
^{\prime }+\mathcal{D}^{\prime }y^{2}\right) yze_{\left( 1\right) i}\right\}
\notag \\
\mathcal{A}^{\prime } &=&\left( \frac{Ze}{4\pi r^{2}}\right) ^{2}\left\{ 
\frac{1}{2}\left( \mathcal{L}_{\mathfrak{FF}}-\mathcal{L}_{\mathfrak{GG}%
}-B^{2}\mathcal{L}_{\mathfrak{FGG}}\right) \right.  \notag \\
&+&\left. \frac{r}{5R}\left( 2\mathcal{L}_{\mathfrak{GG}}-3\mathcal{L}_{%
\mathfrak{FF}}+B^{2}\mathcal{L}_{\mathfrak{FGG}}\right) +\frac{9R}{35r}B^{2}%
\mathcal{L}_{\mathfrak{FGG}}\right\}  \notag \\
\mathcal{B}^{\prime } &=&\left( \frac{Ze}{4\pi r^{2}}\right) ^{2}\frac{R^{2}%
}{r^{2}}\left( 1-\frac{9R}{7r}\right) B^{2}\mathcal{L}_{\mathfrak{FGG}}\,, 
\notag \\
\mathcal{D}^{\prime } &=&-3\left( \frac{Ze}{4\pi r^{2}}\right) ^{2}\left( 
\frac{R}{r}\right) ^{4}\left( 1-\frac{R}{r}\right) B^{2}\mathcal{L}_{%
\mathfrak{FGG}}\,,  \notag \\
\mathcal{C}^{\prime } &=&\left( \frac{Ze}{4\pi r^{2}}\right) ^{2}\left( 
\frac{R}{r}\right) ^{2}\left\{ 2\left( -\mathcal{L}_{\mathfrak{FF}}+\frac{1}{%
2}\mathcal{L}_{\mathfrak{GG}}+B^{2}\mathcal{L}_{\mathfrak{FGG}}\right)
\right.  \notag \\
&+&\left. \frac{3r}{5R}\left( 3\mathcal{L}_{\mathfrak{FF}}-2\mathcal{L}_{%
\mathfrak{GG}}-B^{2}\mathcal{L}_{\mathfrak{FGG}}\right) -\frac{9R}{7r}B^{2}%
\mathcal{L}_{\mathfrak{FGG}}\right\}  \label{out1}
\end{eqnarray}%
where $y=x_{3}/R$ and $z=x_{1}/R$. The ratio $h_{3}^{\mathrm{out}}\left( 
\mathbf{x}\right) /h_{1}^{\mathrm{out}}\left( \mathbf{x}\right) $ can be
expressed as%
\begin{eqnarray}
\frac{h_{3}^{\mathrm{out}}\left( \mathbf{x}\right) }{h_{1}^{\mathrm{out}%
}\left( \mathbf{x}\right) } &=&\frac{\beta ^{\prime }\left( y,z\right) }{yz}%
+\gamma ^{\prime }\left( y,z\right) \frac{y}{z}+E^{\prime }\left( y,z\right) 
\frac{y^{3}}{z}\,,  \notag \\
\beta ^{\prime }\left( y,z\right) &=&\frac{\mathcal{A}^{\prime }}{\mathcal{C}%
^{\prime }+\mathcal{D}^{\prime }y^{2}}\,,\ \ \gamma ^{\prime }\left(
y,z\right) =\frac{\mathcal{B}^{\prime }+\mathcal{C}^{\prime }}{\mathcal{C}%
^{\prime }+\mathcal{D}^{\prime }y^{2}}\,,\ \ E^{\prime }\left( y,z\right) =%
\frac{\mathcal{D}^{\prime }}{\mathcal{C}^{\prime }+\mathcal{D}^{\prime }y^{2}%
}\,,  \label{out2}
\end{eqnarray}%
Taking into account the asymptotic behavior at large magnetic field $B\gg
m^{2}/e$ (\ref{numfactors}), the coefficients $\mathcal{A}^{\prime }$, $%
\mathcal{B}^{\prime }$, $\mathcal{D}^{\prime }$ and $\mathcal{C}^{\prime }$
are

\begin{eqnarray}
&&\mathcal{A}^{\prime }=\left( \frac{Ze}{4\pi r^{2}}\right) ^{2}\left( \frac{%
\alpha }{3\pi B}\right) \left( \frac{e}{m^{2}}\right) \left( \frac{x}{5R}%
\right) \left( 1-\frac{9R^{2}}{7r^{2}}\right) \,,  \notag \\
&&\mathcal{B}^{\prime }=-\left( \frac{Ze}{4\pi r^{2}}\right) ^{2}\left( 
\frac{\alpha }{3\pi B}\right) \left( \frac{e}{m^{2}}\right) \left( \frac{%
R^{2}}{r^{2}}\right) \left( 1-\frac{9R}{7r}\right) \,,  \notag \\
&&\mathcal{D}^{\prime }=3\left( \frac{Ze}{4\pi r^{2}}\right) ^{2}\left( 
\frac{\alpha }{3\pi B}\right) \left( \frac{e}{m^{2}}\right) \left( \frac{R}{r%
}\right) ^{4}\left( 1-\frac{R}{r}\right) \,,  \notag \\
&&\mathcal{C}^{\prime }=\left( \frac{Ze}{4\pi r^{2}}\right) ^{2}\left( \frac{%
\alpha }{3\pi B}\right) \left( \frac{e}{m^{2}}\right) \left( \frac{R}{r}%
\right) ^{2}\left( -1-\frac{3r}{5R}+\frac{9R}{7r}\right) \,,  \label{out3}
\end{eqnarray}%
then $\beta ^{\prime }$, $\gamma ^{\prime },$ and $E^{\prime }$ read%
\begin{eqnarray}
&&\beta ^{\prime }\left( y,z\right) =-\frac{r^{3}}{5MR^{3}}\left( 1-\frac{%
9R^{2}}{7r^{2}}\right) \,,  \notag \\
&&\gamma ^{\prime }\left( y,z\right) =\frac{1}{M}\left( 2+\frac{3r}{5R}-%
\frac{18R}{7r}\right) \,,  \notag \\
&&E^{\prime }\left( y,z\right) =-\frac{3R^{2}}{Mr^{2}}\left( 1-\frac{R}{r}%
\right) \,,  \notag \\
&&M=1+\frac{3r}{5R}-\frac{9R}{7r}-3R^{2}\left( \frac{y}{r}\right) ^{2}\left(
1-\frac{R}{r}\right) \,.  \label{out4}
\end{eqnarray}

\begin{figure}[!th]
\begin{center}
\includegraphics[scale=0.6]{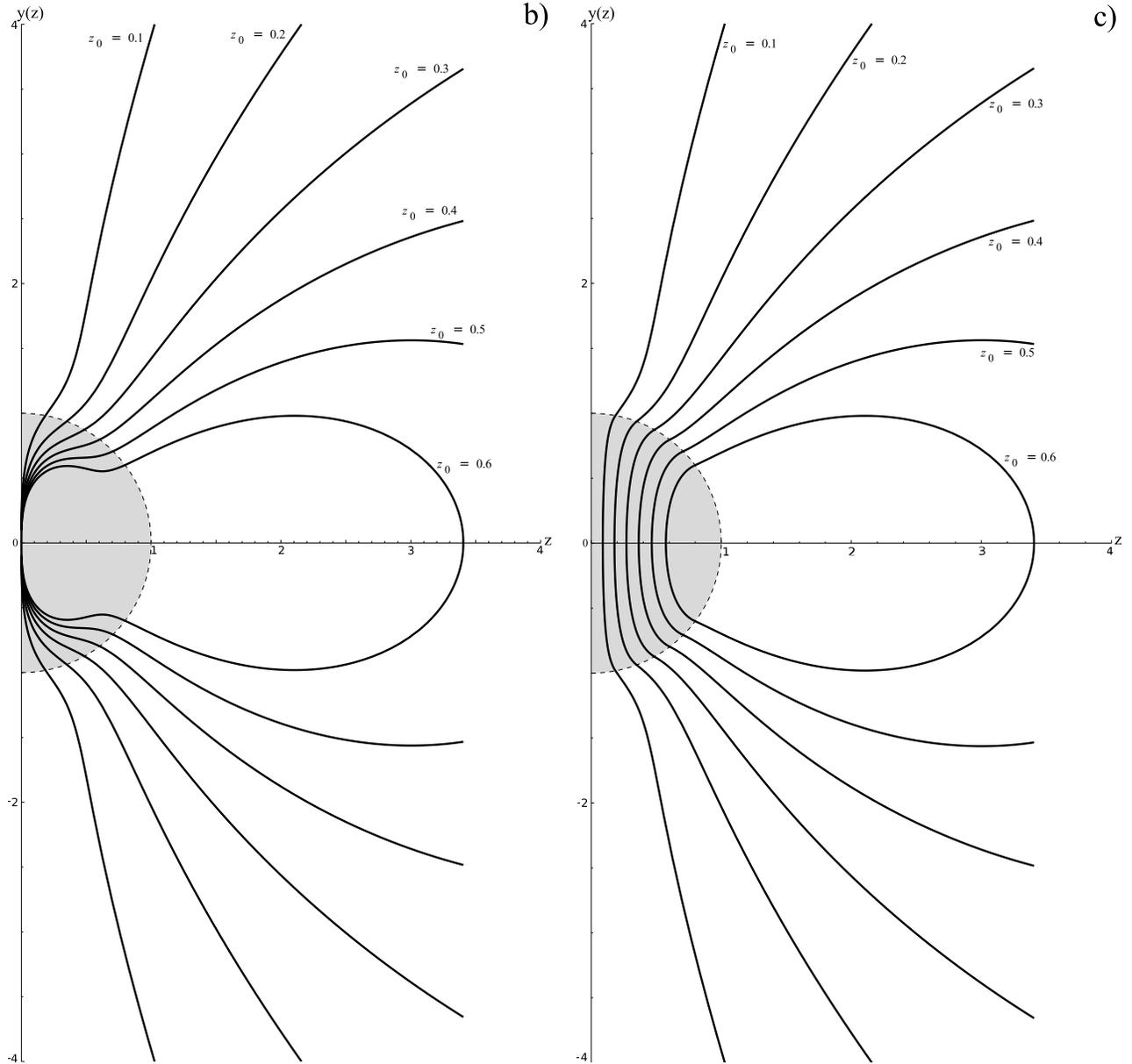}
\end{center}
\caption{{\protect\small {Magnetic dipole lines of a static charge in
external magnetic field exampled with $B=\infty$. b) The pattern to be
trusted outside the charge, $y^{2}+z^{2}>1$, following solution (\protect\ref%
{out5}). For each choice of $z_0$ we extract from (\protect\ref{patinside})
the corresponding value for $y(z)$ at the border of the sphere. See (\protect
\ref{constout}) for some boundary conditions to (\protect\ref{out5}).} c)
United pattern to be trusted throughout.}}
\label{Figoutside}
\end{figure}

Equating the derivative $dy/dz$ with the ratio $h_{3}^{\mathrm{out}}\left( 
\mathbf{x}\right) /h_{1}^{\mathrm{out}}\left( \mathbf{x}\right) $ (\ref{out2}%
) one finds the differential equation for the magnetic lines outside the
sphere. Using (\ref{out2})-(\ref{out4}) the differential equation has the
final form%
\begin{equation}
\frac{dy}{dz}=\frac{9R^{2}r^{4}-7r^{6}+\left( Rry\right) ^{2}\left(
21r^{2}+70Rr-90R^{2}\right) +105R^{5}y^{4}(R-r)}{R^{2}yz\left[ r^{2}\left(
21r^{2}+35Rr-45R^{2}\right) +105R^{3}y^{2}\left( R-r\right) \right] }\,.
\label{out5}
\end{equation}%
This equation does not have analytic (closed) solutions. We found them by
using numerical methods. The integration constant is fixed by the matching
requirements with the pattern in Fig. \ref{Figinside}: we demand that
solutions of (\ref{Solut}), for each fixed $z_{0}$, have the same numerical
values as (\ref{out5}) at the border of the sphere $%
y^{2}+z^{2}=r^{2}/R^{2}=1 $. In this way the continuous continuation of
solutions of (\ref{Solut}) to the outer region, where equation (\ref{out5})
is actual, is achieved. Figure \ref{Figoutside}-c) shows the overall pattern
of magnetic lines to be trusted everywhere, wherefrom the lines beyond their
domains of definition have been deleted. The values of $y\left( z\right) $
and $z$ at the border of the sphere are listed below for each integration
constant $z_{0}$:%
\begin{eqnarray}
z_{0} &=&0.1\rightarrow z\simeq 0.186\,,\ \ y\left( 0.186\right) \simeq
0.983\,,  \notag \\
z_{0} &=&0.2\rightarrow z\simeq 0.349\,,\ \ y\left( 0.349\right) \simeq
0.973\,,  \notag \\
z_{0} &=&0.3\rightarrow z\simeq 0.486\,,\ \ y\left( 0.486\right) \simeq
0.874\,,  \notag \\
z_{0} &=&0.4\rightarrow z\simeq 0.604\,,\ \ y\left( 0.604\right) \simeq
0.797\,,  \notag \\
z_{0} &=&0.5\rightarrow z\simeq 0.707\,,\ \ y\left( 0.707\right) \simeq
0.707\,  \notag \\
z_{0} &=&0.6\rightarrow z\simeq 0.8\,,\ \ y\left( 0.8\right) \simeq 0.6\,.
\label{constout}
\end{eqnarray}

The magnetic lines in \ref{Figoutside}-c) remind very much the standard
pattern of those of a finite-thickness solenoid in classical magnetostatics.

\section{Beyond the spherical symmetry of the applied field}

Here we search for an extension of (\ref{18.9}) to spherically nonsymmetric
applied electric field. Such generalization provides a more general form of
the magnetic dipole moment $\boldsymbol{\mu }$.

Let us first see, how the result (\ref{18.9}), (\ref{18.10}) can be directly
reproduced by considering long-range behavior of the magnetic response (\ref%
{magfield}) to the spherically symmetric electric field (\ref{E}), (\ref{E1}%
). According to (\ref{hgerman}) the field $\mathfrak{h}_{i}\left( \mathbf{x}%
\right) $ in the far-off domain reads 
\begin{equation}
\mathfrak{h}_{i}\left( \mathbf{x}\right) \simeq \frac{1}{2r^{4}}\left( \frac{%
Ze}{4\pi }\right) ^{2}\left[ \left( \mathcal{L}_{\mathfrak{FF}}-\mathcal{L}_{%
\mathfrak{FGG}}\frac{\left( \mathbf{B\cdot x}\right) ^{2}}{r^{2}}\right)
B_{i}-2\mathcal{L}_{\mathfrak{GG}}\frac{\left( \mathbf{B\cdot x}\right) x_{i}%
}{r^{2}}\right] \,,  \label{19.3}
\end{equation}%
where we have restricted ourselves to the leading contribution at large $r=|%
\mathbf{x}|$. The leading behavior of the quantities $\mathfrak{I}_{k}\left( 
\mathbf{x}\right) $ in (\ref{magfield}) is,%
\begin{equation}
\mathfrak{I}_{k}(r)\simeq \frac{1}{r}\int d^{3}y\mathfrak{h}_{k}(\mathbf{y}),
\notag
\end{equation}%
provided that the integrals here converge. Then,%
\begin{equation}
\partial _{i}\partial _{k}\mathfrak{I}_{k}(\mathbf{x})=\left( \frac{%
3x_{i}x_{k}}{r^{5}}-\frac{\delta _{ik}}{r^{3}}\right) \int d^{3}y\mathfrak{h}%
_{k}(\mathbf{y})\,.  \label{ddJ}
\end{equation}%
The field $\mathfrak{h}_{i}\left( \mathbf{x}\right) $ (\ref{19.3}) falls off
faster than this, namely as $1/r^{4},$ hence its contribution into the first
line of (\ref{magfield}) can be neglected as compared to (\ref{ddJ}). So,
the large-distance behavior of the nonlinear magnetic field $h_{i}\left( 
\mathbf{x}\right) $ (\ref{magfield}) is just (\ref{ddJ}), i.e., that of a
magnetic dipole. Its magnetic dipole moment $\mu _{i}^{\mathrm{LD}}$ is%
\begin{eqnarray}
&&\mu _{i}^{\mathrm{LD}}=\frac{1}{4\pi }\int d^{3}y\mathfrak{h}_{i}(\mathbf{y%
})\,,  \label{20} \\
&&\mathfrak{h}_{i}(\mathbf{y})=\frac{B_{i}}{2}\left( \mathcal{L}_{\mathfrak{%
FF}}\mathbf{E}^{2}\left( \mathbf{y}\right) -\mathcal{L}_{\mathfrak{FGG}%
}\left( \mathbf{B}\cdot \mathbf{E}\left( \mathbf{y}\right) \right)
^{2}\right) -\mathcal{L}_{\mathfrak{GG}}\left( \mathbf{B}\cdot \mathbf{E}%
\left( \mathbf{y}\right) \right) E_{i}\left( \mathbf{y}\right) \,.  \notag
\end{eqnarray}%
The result (\ref{20}) agrees with the previous result (\ref{18.10}) in case
the spherically symmetric $\mathbf{E}$\ is specialized to (\ref{1.1}). To
make sure of this it suffices to substitute expression (\ref{19a}) for $%
\mathfrak{h}_{i}(\mathbf{y})$ into (\ref{20}) and fulfill the integration,
which converges both at the lower, $y=0,$ and the upper, $y=\infty ,$ limit.

However, the validity of the result (\ref{20}) is much wider. For instance,
let us take Eq. (\ref{Coulomb}) or, equivalently, (\ref{Coulomb1}) for the
scalar potential responsible for the remote cylindrically-symmetric electric
field of a static extended charge $Q=Ze,$ whose density decreases
sufficiently fast at infinity, but is otherwise arbitrary, not subject to
any symmetry. Recall, that this cylindrical, instead of spherical, symmetry
became in Sec. 2.1 the effect of the linear vacuum polarization in an
external magnetic field. It is easy to make sure that when this electric
field is substituted into (\ref{hgerman}) for $\mathfrak{h}_{i}\left( 
\mathbf{x}\right) ,$ the resulting expression in place of (\ref{19.3}) also
decreases as $1/r^{4},$ the same as it. Hence, we are left again with Eq. (%
\ref{ddJ}) \ for the large-distance asymptote of the nonlinearly induced
magnetic field $h_{i}\left( \mathbf{x}\right) $ (\ref{magfield}). Since the
electric field (\ref{Coulomb1}) is invariant under rotations around the
external magnetic field $\mathbf{B,}$ the latter remains the only special
direction in the space. Consequently, the magnetic moment (\ref{20}) is
directed along $\mathbf{B,}$ the same as (\ref{18.10}).

\section{Conclusion}

In this work it was shown that a static charge, apart from being a source of
the customary Coulomb-like electric field, is also a nonlinear source of a
magnetic field. This field is generated due to a nonlinear induced current
caused by a constant and homogeneous external magnetic field. As a result,
the long-range magnetic field behaves like a magnetic field generated by a
magnetic dipole moment. In other words, the extended charge has a long-range
magnetic dipole character. The magnetic field lines resemble the well known
magnetic dipole structure.

The validity of equations found here for the nonlinear magnetic response of
the magnetic background to an applied electric field is restricted to the
fields, smooth in time and space. They can be directly applied to charged
large astrophysical objects, but lead to overestimation, where small objects
as charged mesons and baryons are concerned. Therefore, to make such
application reasonable, one needs to go beyond the infrared approximation.
To this end QED calculations of three-photon diagram in an external magnetic
field must be efficiently exploited beyond the photon mass shell. We hope to
come back to this more complicated problem in future works.

\section*{Acknowledgements}

T. C. Adorno acknowledges the financial support of FAPESP under the process
2013/00840-9 and 2013/16592-4. He is also thankfull for the Department of
Physics of the University of Florida for the kind hospitality. D. Gitman
thanks CNPq and FAPESP for permanent support, in addition his work is done
partially under the project 2.3684.2011 of Tomsk State University. A. Shabad
acknowledges the support of FAPESP, Processo 2011/51867-9, and of RFBR under
the Project 11-02-00685-a. He also thanks USP for kind hospitality extended
to him during his stay in Sao Paulo, Brazil, where this work was partially
fulfilled. The authors are thankful to C. Costa for discussions.

\section*{Appendix}

Performing derivations of the potential (\ref{1.1}) leads to Dirac
delta-functions and their derivatives as well. Taking into account
smoothness conditions at $r=R$, one is able to simplify the explicit form of
some quantities under consideration. One can see that,%
\begin{equation}
a_{0}^{\mathrm{I}}\left( R\right) =a_{0}^{\mathrm{II}}\left( R\right) \,,\ \
\left. \frac{da_{0}^{\mathrm{I}}\left( r\right) }{dr}\right\vert
_{r=R}=\left. \frac{da_{0}^{\mathrm{II}}\left( r\right) }{dr}\right\vert
_{r=R}\,,\ \ \left. \frac{d^{2}a_{0}^{\mathrm{I}}\left( r\right) }{dr^{2}}%
\right\vert _{r=R}\neq \left. \frac{d^{2}a_{0}^{\mathrm{II}}\left( r\right) 
}{dr^{2}}\right\vert _{r=R}\,,  \label{app0.1}
\end{equation}%
\newline
and higher derivatives are not continuous at $r=R$. In this way any function
proportional to $da_{0}\left( r\right) /dr$,%
\begin{equation}
\frac{da_{0}\left( r\right) }{dr}=\frac{da_{0}^{\mathrm{I}}\left( r\right) }{%
dr}\theta \left( R-r\right) +\frac{da_{0}^{\mathrm{II}}\left( r\right) }{dr}%
\theta \left( r-R\right) +\left( a_{0}^{\mathrm{II}}\left( r\right) -a_{0}^{%
\mathrm{I}}\left( r\right) \right) \delta \left( r-R\right) \,,
\label{app0.2}
\end{equation}%
can be simplified \ by omitting the Dirac delta-function terms. This
simplification is supported by the fact that $\left( a_{0}^{\mathrm{II}%
}\left( r\right) -a_{0}^{\mathrm{I}}\left( r\right) \right) \delta \left(
r-R\right) $ gives zero contribution, since%
\begin{equation*}
\int_{-\infty }^{\infty }drf\left( r\right) \left( a_{0}^{\mathrm{II}}\left(
r\right) -a_{0}^{\mathrm{I}}\left( r\right) \right) \delta \left( r-R\right)
=f\left( R\right) \left( a_{0}^{\mathrm{II}}\left( R\right) -a_{0}^{\mathrm{I%
}}\left( R\right) \right) =0\,,
\end{equation*}%
where $f\left( r\right) $ represents any function, well-behaved at $r=R$.
The same idea can be generalized to any function which depends on $%
d^{2}a_{0}\left( r\right) /dr^{2}$ or higher derivatives.

In order to evaluate the integrals $\mathfrak{I}_{k}\left( \mathbf{x}\right) 
$, which make part of the total magnetic field (\ref{magfield}), one has to
evaluate $v\left( r\right) ,\ u\left( r\right) $ and $w\left( r\right) $ (%
\ref{v})-(\ref{U}). All of these functions can be conveniently written as
sums of two other integrals. For example, we write (\ref{v}) as $v\left(
r\right) =\left( 2\pi /r\right) \left[ v_{1}\left( r\right) +v_{2}\left(
r\right) \right] $ where,%
\begin{equation*}
v_{1}\left( r\right) =\int_{0}^{R}dy\left[ \left( r+y-\left\vert
r-y\right\vert \right) \frac{y^{3}}{R^{6}}\right] \,,\ \ v_{2}\left(
r\right) =\int_{R}^{\infty }dy\left( \frac{r+y-\left\vert r-y\right\vert }{%
y^{3}}\right) \,.
\end{equation*}%
Now, $v_{1}\left( r\right) $ can be calculated considering two situations,
namely $r<R$ and $r>R$. Then%
\begin{eqnarray*}
v_{1}\left( r\right) &=&\frac{1}{R^{6}}\left[ \int_{0}^{r}dyy^{3}\left(
2y\right) +\int_{r}^{R}dyy^{3}\left( 2r\right) \right] =\frac{r}{2R}\left( 1-%
\frac{r^{4}}{5R^{4}}\right) \,,\ \ r<R\,, \\
v_{1}\left( r\right) &=&\frac{2}{R^{6}}\int_{0}^{R}dyy^{4}=\frac{2}{5R}\,,\
\ r>R\,,
\end{eqnarray*}%
hence,%
\begin{equation}
v_{1}\left( r\right) =\frac{r}{2R}\left( 1-\frac{r^{4}}{5R^{4}}\right)
\theta \left( R-r\right) +\frac{2}{5R}\theta \left( r-R\right) \,.
\label{app1}
\end{equation}%
Similarly%
\begin{equation}
v_{2}\left( r\right) =\frac{r}{R^{2}}\theta \left( R-r\right) +\left( \frac{2%
}{R}-\frac{1}{r}\right) \theta \left( r-R\right) \,.  \label{app2}
\end{equation}%
Then (\ref{v}) takes the final form (\ref{6a}).

Besides, it should be noted that $u\left( r\right) $ and $w\left( r\right) $
can be written in a simplified form,%
\begin{eqnarray}
x^{2}u\left( r\right) &=&\frac{1}{2}\left( \frac{3c\left( r\right) }{r^{2}}%
-v\left( r\right) \right) \,,\ \ w(r)=\frac{1}{2}\left( v\left( r\right) -%
\frac{c\left( r\right) }{r^{2}}\right) \,,  \notag \\
c\left( r\right) &=&2\pi r^{2}\int_{0}^{\infty }dyy^{4}\mathcal{E}^{2}\left(
y\right) V_{2}\left( r,y\right) \,,  \label{app4.0}
\end{eqnarray}%
such that after finding $c\left( r\right) $ we can immediately derive $%
u\left( r\right) $ and $w\left( r\right) $. Thus, using the angular integral
(\ref{V.2}), the function $c\left( r\right) $ takes the form%
\begin{equation}
c\left( r\right) =\frac{4\pi }{15r}\left[ \int_{0}^{r}dy\left(
5r^{2}y^{4}+2y^{6}\right) \mathcal{E}^{2}\left( y\right) +\int_{r}^{\infty
}dy\left( 2r^{5}y+5r^{3}y^{3}\right) \mathcal{E}^{2}\left( y\right) \right]
\,.  \label{app5.1}
\end{equation}%
Considering again $r<R$ and $r>R$, separately, we list below each integral
appearing above:%
\begin{eqnarray*}
&&\int_{0}^{r}dy\left( 5r^{2}y^{4}+2y^{6}\right) \frac{\theta \left(
R-y\right) }{R^{6}}=\frac{9r^{7}}{7R^{6}}\theta \left( R-r\right) +\left( 
\frac{7r^{2}+2R^{2}}{7R}\right) \theta \left( r-R\right) \,; \\
&&\int_{0}^{r}dy\left( 5r^{2}y^{4}+2y^{6}\right) \frac{\theta \left(
y-R\right) }{y^{6}}=\left( \frac{5r^{2}-3rR-2R^{2}}{R}\right) \theta \left(
r-R\right) \,; \\
&&\int_{r}^{\infty }dy\left( 2r^{5}y+5r^{3}y^{3}\right) \frac{\theta \left(
R-y\right) }{R^{6}}=\left( \frac{5r^{3}}{4R^{2}}+\frac{r^{5}}{R^{4}}-\frac{%
9r^{7}}{4R^{6}}\right) \theta \left( R-r\right) \,; \\
&&\int_{r}^{\infty }dy\left( 2r^{5}y+5r^{3}y^{3}\right) \frac{\theta \left(
y-R\right) }{y^{6}}=\frac{r^{3}}{R^{2}}\left( \frac{r^{2}}{2R^{2}}+\frac{5}{2%
}\right) \theta \left( R-r\right) +3r\theta \left( r-R\right) \,.
\end{eqnarray*}%
Substituting these results in (\ref{app5.1}) and using (\ref{app4.0}), the
scalar functions take their final form (\ref{6b}).

\end{document}